\documentclass[titlepage,a4paper,12pt,feynmf]{article}
\usepackage[affil-it]{authblk}
\usepackage[english]{babel}
\usepackage{blindtext}
\usepackage{amssymb}
\usepackage{amsmath,amssymb,mathrsfs}
\usepackage{graphicx}
\usepackage{ifpdf}
\usepackage{xcolor}
\newcommand*{\rom}[1]{\expandafter\@slowromancap\romannumeral #1@}
\renewcommand{\thefootnote}{\fnsymbol{footnote}}
\title{Transverse expansion of hot magnetized Bjorken flow  in heavy ion collisions}

\author[1,2]{M. Haddadi Moghaddam}
\author[1]{B. Azadegan}
\author[1]{A. F. Kord \thanks{corresponding author: A. F. Kord \\
		Email address:afarzaneh@hsu.ac.ir}}

\author[2]{W. M. Alberico}
\affil[1]{Department of Physics, Hakim Sabzevari University (HSU), P.O.Box 397, Sabzevar, Iran}
\affil[2]{ Department of Physics, University of Turin and INFN, Turin, Via P. Giuria 1, I-10125 Turin, Italy}
\date{}

\begin{document}
	\def\ctr#1{\hfil $\,\,\,#1\,\,\,$ \hfil}
	\def\tstrut{\vrule height 2.7ex depth 1.0ex width 0pt}
	\def\mystrut{\vrule height 3.7ex depth 1.6ex width 0pt}
	\def \inparg{\leftskip = 40pt\rightskip = 40pt}
	\def \outparg{\leftskip = 0 pt\rightskip = 0pt}
	\def\lf{16\pi^2}
	\def\beqn{\begin{eqnarray}}
		\def\eeqn{\end{eqnarray}}
	\def\sy{supersymmetry}
	\def\sic{supersymmetric}
	\def\sa{supergravity}
	\def\ssm{supersymmetric standard model}
	\def\sm{standard model}
	\def\ssb{spontaneous symmetry breaking}
	\def\smgroup{$SU_3\otimes\ SU_2\otimes\ U_1$}
	\def\app{{Acta Phys.\ Pol.\ }{\bf B}}
	\def\anp{Ann.\ Phys.\ }
	\def\cmp{Comm.\ Math.\ Phys.\ }
	\def\fortphys{{Fort.\ Phys.\ }{\bf A}}
	\def\ijmpa{{Int.\ J.\ Mod.\ Phys.\ }{\bf A}}
	\def\jetp{JETP\ }
	\def\jetpl{JETP Lett.\ }
	\def\jmp{J.\ Math.\ Phys.\ }
	\def\mpla{{Mod.\ Phys.\ Lett.\ }{\bf A}}
	\def\nc{Nuovo Cimento\ }
	\def\npb{{Nucl.\ Phys.\ }{\bf B}}
	\def\physrep{Phys.\ Reports\ }
	\def\plb{{Phys.\ Lett.\ }{\bf B}}
	\def\pnas{Proc.\ Natl.\ Acad.\ Sci.\ (U.S.)\ }
	\def\pr{Phys.\ Rev.\ }
	\def\prd{{Phys.\ Rev.\ }{\bf D}}
	\def\prl{Phys.\ Rev.\ Lett.\ }
	\def\ptp{Prog.\ Th.\ Phys.\ }
	\def\sjnp{Sov.\ J.\ Nucl.\ Phys.\ }
	\def\tmp{Theor.\ Math.\ Phys.\ }
	\def\pw{Part.\ World\ }
	\def\zpc{Z.\ Phys.\ {\bf C}}
	\def\pa{\partial}
	%\author[1]{Darth Vader\thanks{corresponding author}}
	
	\maketitle
	%\blindtext
	\renewcommand{\thefootnote}{\arabic{footnote}}
	%\vskip .3in {\textbf{\Large{ Transverse expansion of hot magnetized Bjorken flow  in heavy ion collisions}}}
	%\medskip
	%\vskip .3in \centerline{ \bf  M. Haddadi Moghaddam$^{1,2}$,  B. Azadegan   $^1$,  A. F. Kord$^{1}$, W. M. Alberico$^3$ }
	%\bigskip
	%{\small{\it {$^1$: Department of Physics, Hakim Sabzevari University (HSU), P.O.Box 397, Sabzevar, Iran\\
	% $^2$: Institute for Research  in Fundamental Sciences(IPM),P.O. Box 19395-5531, Tehran, Iran\\
	%$^2$: Department of Physics, University of Turin and INFN, Turin, Via P. Giuria 1, I-10125 Turin, Italy }}}
	
	%\vskip .001in

\begin{abstract}
 We argue that the existence of an inhomogeneous external magnetic field can lead to radial flow
in transverse plane. Our aim is to show how the introduction of a magnetic field generalizes the Bjorken flow.
We investigate the effect of an inhomogeneous weak external magnetic field on the transverse
expansion of in-viscid fluid created in high energy nuclear collisions. In order to simplify our
calculation and compare with Gubser model, we consider the fluid under investigation to be produced
in central collisions, at small impact parameter; azimuthal symmetry has been considered.
In our model, we assume an inhomogeneous external magnetic field following the power-law decay in proper
time and having radial inhomogeneity perpendicular to the radial velocity of the in-viscid fluid in the
transverse plane; then the space time evolution of the transverse expansion
of the fluid is obtained. We also show how the existence of an inhomogeneous external magnetic
field modifies the energy density.
Finally we use the solutions for the transverse velocity and energy density
in the presence of a weak magnetic field, to estimate the transverse momentum spectrum of protons and pions
emerging from the Magneto-hydrodynamic solutions.
\end{abstract}


%**********************************************************************************
\section{Introduction}
 Collisions of two heavy nuclei at high energy produce a hot and dense fireball.
Quarks and gluons could reach the deconfined state, called quark gluon plasma (QGP),
in a very short time ($\sim$ 1fm/c) after the initial hard parton collisions of nuclei.
A very handy model which describes the typical motion of partons after collision is the
Bjorken flow model \cite{bj83}.
 Based on some assumptions such as boost invariance along beam line, translation and rotation
invariance in the transverse plane, one can show that all quantities of interest only depend on the
proper time $\tau$ and not on the transverse ($x_{\perp}, \phi$) coordinates, nor
on the rapidity $\eta$.
 Using the above assumptions, together with invariance under reflection $\eta \rightarrow -\eta$,
one can determine the four-velocity profile.  The four-velocity is  $ u^{\mu}=(1,0,0,0) $
in the $(\tau, x_{\perp},\phi,\eta)$ coordinate system.  Besides, it is straightforward to show that the
energy density decays as ${\tau^{-4/3}}$ in the local rest frame if the medium is equilibrated and the
equation of state of the medium is $p=\epsilon/3$.

Based on the size of the colliding nuclei, one realizes that assuming translational invariance
in the transverse plane is not realistic \cite{gubser}.
Using the Bjorken model one often assumes that in medium the radial flow ($u_{\perp}$) is zero.
However, this assumption is not correct even for central collisions, and it might mislead the
subsequent hydrodynamical flow, on which much of heavy-ions phenomenology depends.

The aim of our work is to generalize the Bjorken model by considering an inhomogeneous external magnetic
field acting on the medium. We show that the presence of the magnetic field leads to non-zero radial flow.
In order to simplify our calculation, we consider central heavy ions collisions. We still consider
rotational symmetry around beam line, as well as boost invariance  along the beam line.
However, we assume that translational invariance in the transverse plane is broken by the magnetic field.
Then we obtain a four-velocity profile which has a non zero radial component.
In the present study for central collisions (small impact parameter), we provide an analytical solution
for the transverse expansion of a hot magnetized plasma, based on perturbation theory.

We concentrate on the special case of a (2+1) dimensional, longitudinally boost-invariant fluid expansion
as the Bjorken flow; the fluid also radially expands in the transverse plane, under the influence of
an inhomogeneous external magnetic field which is transverse to
the radial fluid velocity (this proceeds according to the so called{ \it{transverse}} MHD).

We consider an inviscid fluid coupled to an external magnetic field. As one expects in central collisions,
we assume that the external magnetic fields is small compared to the fluid energy density \cite{pu15}.
Therefore, we can neglect the coupling to the Maxwell's equations and solve the conservation equations
perturbatively and analytically \cite{pu}.

Moreover the presence of external magnetic field may induce internal electromagnetic fields
of the fluid. The internal magnetic fields are dictated by Maxwell's equations and one should solve the conservation
equations and Maxwell's equations coupled to each other by numerical methods \cite{Haddadi et al}.
In this work we neglect the effects of such internal magnetic field. Hence we will consider the system with an
inhomogeneous external magnetic field and will investigate the anisotropic transverse flow and the modified energy
density of the fluid induced by the external magnetic fields. As in Gubser flow, the
finite size of the colliding nuclei leads to non-zero radial velocity ($u_\perp$); we show that the
inhomogeneous weak external magnetic field also leads to nonzero radial velocity and can produce modifications on
the radial expansion of the plasma in central collisions.

We remind the reader that recently a wide range of studies has shown that relativistic heavy-ion collisions create also huge
magnetic field due to the relativistic motion of the colliding heavy ions carrying large positive
electric charge \cite{Tuchin:2013apa}-\cite{voronyuk}. The interplay of magnetic field and QGP matter
has been predicted to lead to a number of interesting phenomena. One can see recent reviews on this topic in
Refs. \cite{kharzeev2008}-\cite{ kharzeevprl}  for more details.

Previous theoretical studies show that the strength of the produced magnetic field depends on the
center of mass energy ($\sqrt{s_{NN}}$) of the colliding nuclei, on the impact parameter ($b$)
of the collision, on the electrical and chiral conductivities ($\sigma_{el}, \sigma_{\chi}$)
of the medium \cite{Tuchin:2013ie, Deng:2012pc, chinees, a15}.
 Moreover, the magnetic field in central collisions becomes non-zero due to the fluctuating proton
position from event to event \cite{{pu15},{Bzdak:2011yy}}.
 It has been found that the ratio of magnetic field energy to the fluid energy density
($\sigma=e B^2/2\epsilon$) in central collisions
 is much smaller than in peripheral collisions \cite{pu15}.
The authors of Ref. \cite{pu15} computed the fluid energy density and electromagnetic field by
using the Monte Carlo Glauber model.
The initial energy density for the fluid at proper time $\tau_{i}=0.5$ fm was fixed to
$\sim 40 $~GeV/fm$^{3}$. They found $\sigma\ll1$ for most of the events,
 at the center of the collision zone and for impact parameter $b=0$, while for large $b$ as
compared to central collisions, $\sigma$ becomes larger as a result of the increase in magnetic
field and decrease in fluid energy density.
 In a plasma $\sigma=1$ indicates that the effect of magnetic field in the plasma evolution can not be neglected,
 but it is worth observing that in some situations, even $\sigma\sim0.01$ may affect
 the hydrodynamical evolution \cite{pu15}.

Recently, some efforts in numerical and analytical works have been made, based on the relativistic
magneto-hydrodynamic  (RMHD) setup, to describe
high energy heavy ion collisions (See, for example, \cite{pu} and \cite{roy15}-\cite{hattori}).
In \cite{pu} the goal was to obtain an analytical solution in (1+1) dimensional Bjorken flow
for ideal transverse RMHD, and the conservation equations were solved perturbatively and analytically.
In our previous work \cite{Haddadi et al},
we developed a simple code for transverse expansion in (1+1)D RMHD setup
in order to solve coupled conservation equations and Maxwell's equations numerically.

We found that this coupling can indeed affect the solutions with respect to the ones of ref.\cite{pu}.
In the present work, we show that the perturbative approach of Ref.\cite{pu}
can be applied to the case of central collision,
in order to find analytical solution for the transverse expansion of QGP matter
in the presence of an  external magnetic field.

The paper is organized as follows. In Section $2$,  we introduce the ideal relativistic
magnetohydrodynamic equations in their most general form,
considering them in the case of a plasma with infinite electrical conductivity.
In Section $3$  we present our perturbative approach and the analytical solutions we found.
Section $4$ illustrates and discusses the general results obtained.
Section $5$ contains a calculation of the transverse momentum spectrum together woth a comparison
of this quantity with experimental results obtained at RHIC.
 Conclusions and subsequent outlook can be found in the last section.

%***************************************************************************************
 \section{Ideal Relativistic magneto-hydrodynamic }
We deal with the case of an ideal non-resistive plasma, with vanishing electric field in the local
rest-frame ($e^\mu=0$), which is embedded in an external magnetic field ($b_\mu$)
\cite{geod}-\cite{an89}.
The energy momentum conservation equations read:
\begin{eqnarray}\label{eq:conserve}
d_\mu(T_{pl}^{\mu\nu}+T_{em}^{\mu\nu})=0,
\end{eqnarray}
where
\begin{eqnarray}
T_{pl}^{\mu\nu}&=&(\epsilon+P)u^\mu u^\nu+Pg^{\mu\nu}\\
T_{em}^{\mu\nu}&=&b^2 u^\mu u^\nu+\frac{1}{2}b^2 g^{\mu\nu}-b^\mu b^\nu .
\end{eqnarray}
In the above $g_{\mu\nu}$ is the metric tensor, $\epsilon$ and $P$ are the energy density and
pressure, respectively. Moreover $d_\mu$ is the covariant derivative, defined later in equation (\ref{derivative}).

The four velocity is defined as
$$u_\mu=\gamma(1, \vec v),\ \gamma=\frac{1}{\sqrt{1-v^2}}$$
satisfying the condition $u^\mu u_\mu=-1$.

Canonically one takes projections of the equation $d_\mu(T_{pl}^{\mu\nu}+T_{em}^{\mu\nu})=0$
along the parallel and perpendicular directions to $u_\nu$.
The parallel projection is obtained via $u_\nu d_\mu(T_{pl}^{\mu\nu}+T_{em}^{\mu\nu})$, which gives:
\begin{eqnarray}\label{eq:energycons}
D(\epsilon+b^2/2)+(\epsilon+P+b^2)\Theta+u_\nu b^\mu(d_\mu b^\nu)&=&0,
\end{eqnarray}
For the transverse projection we use the definition $\Delta^{\mu\nu}=g^{\mu\nu}+u^\mu u^\nu$; then
$\Delta^\alpha_\nu d_\mu(T_{pl}^{\mu\nu}+T_{em}^{\mu\nu})=0 $ gives:
\begin{eqnarray}\label{eq:momentumcons}
(\epsilon+P+b^2)Du^\alpha=-\nabla^\alpha(P+\frac{1}{2}b^2)+d_\mu (b^\mu b^\alpha)+u^\alpha
u_\nu d_\mu( b^\mu b^\nu).
\end{eqnarray}
Notice that $\alpha$ should be a spacelike index. Moreover
\begin{eqnarray}
D=u^\mu d_\mu,\ \ \ , \Theta=d_\mu u^\mu,\ \ \ , \nabla^\alpha=\Delta^\alpha_{\ \nu}d^\nu.
\end{eqnarray}

%***************************************************************************
\section{Ideal transverse MHD setup in the transverse expansion}
We assume that the medium has a finite transverse size and expands both radially and
along the beam axis, the only nonzero
components of $u_\mu=(u_\tau, u_\perp, 0, 0)$ being $u_\tau$, which describes the boost invariant
longitudinal expansion, and $u_\perp$, which describes the transverse expansion. For the sake of
simplicity we suppose that $u_\phi=0$ because we claimed that the system is rotationally symmetric.

\begin{figure}[h!]
\begin{center}
\includegraphics[width=2 in]{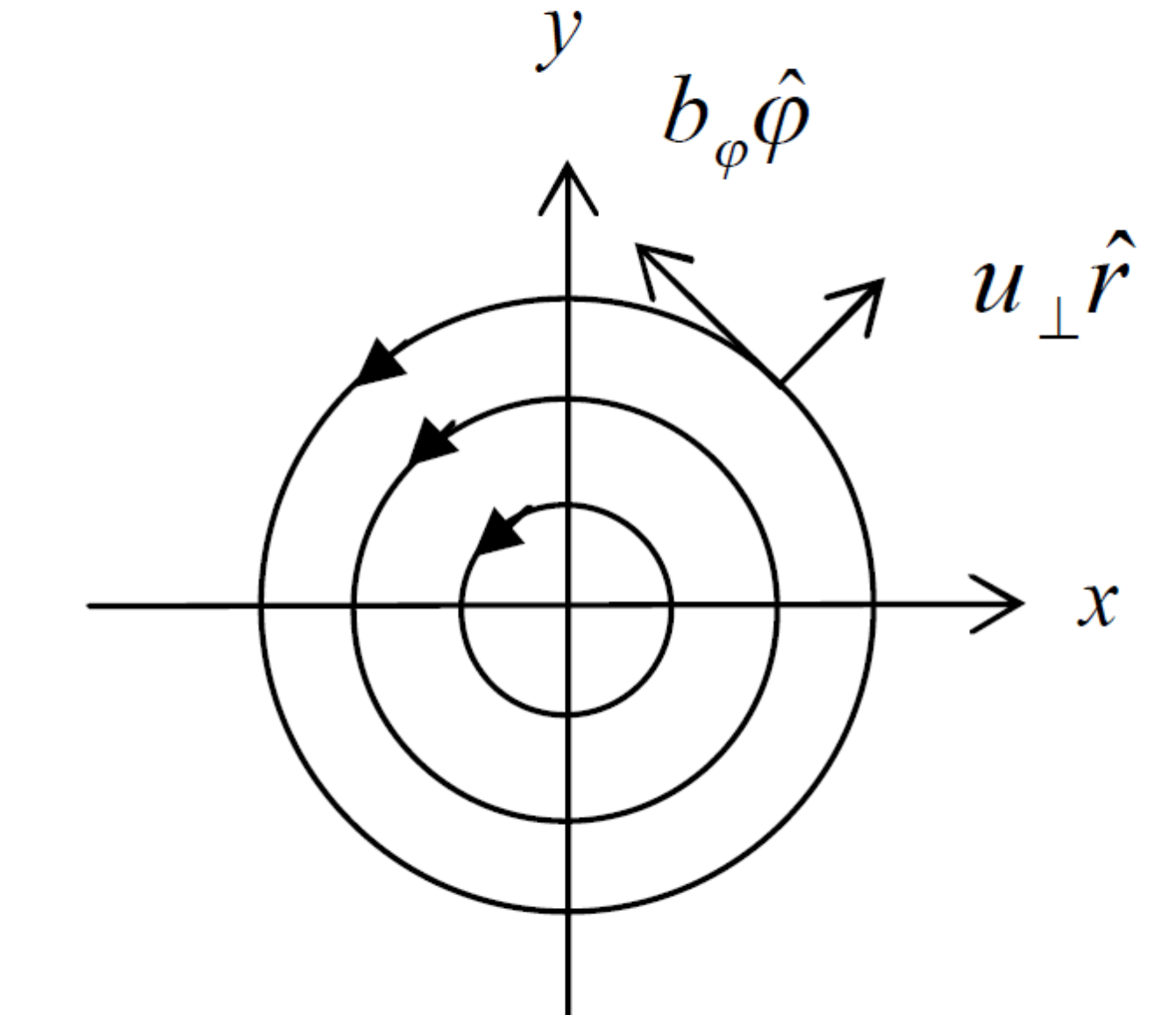} \\
\end{center}
\caption{Transverse MHD $\textbf{u}\cdot\textbf{B}=0$.}
\label{fig:diagram}
\end{figure}

It is more convenient to work in Milne coordinates, $x^m=(\tau, x_\perp, \phi, \eta)$, such that:
\begin{eqnarray}
x=x_\perp\cos\phi,\ y=x_\perp\sin\phi, z=\tau\sinh\eta,\ t=\tau\cosh\eta,\nonumber\\
\tau=\sqrt{t^2-z^2},\ \eta=\frac{1}{2}\ln\frac{t+z}{t-z},\ \phi=\tan^{-1}(y/x),\ x_\perp^2=x^2+y^2
\end{eqnarray}
Moreover we suppose that the external magnetic field is located in the transverse plane as
$b_\mu=(0, 0, b_\phi, 0)$
where $b^\mu b_\mu\equiv b^2$ is defined. Our setup is depicted in Fig. \ref{fig:diagram}.
The metric for the coordinates $(\tau, x_\perp, \phi, \eta)$ is parameterized  as follows:
$g_{\mu\nu}=diag(-1, 1, x_\perp^2, \tau^2)$ and $g^{\mu\nu}=diag(-1, 1,1/ x_\perp^2, 1/\tau^2)$.
Correspondingly
\begin{eqnarray}
ds^2=-d\tau^2+dx_\perp^2+x_\perp^2d\phi^2+\tau^2 d\eta^2.
\end{eqnarray}
In this configuration it is found that $u^\tau=-u_\tau=-u_0$ and $\partial^\tau=-\partial_\tau$.

We have to take care of the following covariant derivative (instead of the usual one):
\begin{eqnarray}
d_\mu A^\mu=\partial_\mu A^\mu+\Gamma^\mu_{\mu\rho}A^\rho,
\label{derivative}
\end{eqnarray}
where the Cristoffel symbols are defined as follows:
\begin{equation}
\Gamma^i_{jk}=\frac{1}{2}g^{im}\Big(\frac{\partial g_{mj}}{\partial x^k}+\frac{\partial g_{mk}}
{\partial x^j}-\frac{\partial g_{jk}}{\partial x^m}\Big).
\end{equation}
Here we frequently take advantage of the following formula:
\begin{eqnarray}
\Gamma^i_{jk}&=&0, \ for\ i\neq j\neq k\\
\Gamma^i_{jj}&=&-\frac{1}{2g_{ii}}\frac{\partial g_{jj}}{\partial x^i},\ for\ i\neq j\\
\Gamma^i_{ij}&=&\Gamma^i_{ji}=\frac{1}{2g_{ii}}\frac{\partial g_{ii}}{\partial x^j}
=\frac{1}{2}\frac{\partial \ln g_{ii}}{\partial x^j}.
\end{eqnarray}
Hence the only non-zero Christoffel symbols, here, are $\Gamma^\tau_{\eta\eta}=\tau,\
\Gamma^{x_\perp}_{\phi\phi}=-x_\perp,\ \Gamma^\phi_{x_\perp\phi}=\frac{1}{x_\perp},\
\Gamma^\eta_{\tau\eta}=\frac{1}{\tau}$.
Now  $D$ and $\Theta$ are given by:
\begin{equation}
 D=-u_0\partial_\tau+u_\perp\partial_\perp,\ \Theta=-\partial_\tau u_0+\frac{u_\perp}{x_\perp}+
\frac{\partial u_\perp}{\partial x_\perp}-\frac{u_0}{\tau}.
\end{equation}
The constraint $u^2=u^\tau u_\tau+u^\perp u_\perp=-u_0^2+u_\perp^2=-1$ must be satisfied as well.

We now look for the perturbative solution of the conservation equations in the presence
of a weak external inhomogeneous magnetic field pointing along the $\phi$ direction
in an inviscid fluid with infinite electrical conductivity and obeying Bjorken flow
in $z$ - direction. Our setup is given by:
\begin{eqnarray}
&&b_\mu=(0, 0, \lambda b_\phi, 0), \ u_\mu=(1, \lambda^2 u_\perp, 0, 0),\\&&
\epsilon=\epsilon_0(\tau)+\lambda^2\epsilon_1(\tau, x_\perp),\ \epsilon_0(\tau)=\frac{\epsilon_c}{\tau^{4/3}},
\end{eqnarray}
where $\epsilon_c$ is the energy density at proper time $\tau_0$. Then the energy conservation and Euler equations
[Eqs. (4), (5)] reduce to two coupled differential equations.
Up to $O(\lambda^2)$, they are:
\begin{eqnarray}
\partial_\tau\epsilon_1-\frac{4\epsilon_c}{3\tau^{4/3}}\left(\frac{u_\perp}{x_\perp}+\frac{\partial
u_\perp}{\partial x_\perp}\right)+\frac{4\epsilon_1}{3\tau}+b_\phi\partial_\tau b_\phi+\frac{b_\phi^2}{\tau}&=&0\label{Euler1}\\
\partial_\perp\epsilon_1-\frac{4\epsilon_c}{\tau^{4/3}}\partial_\tau u_\perp+\frac{4\epsilon_c}
{3\tau^{7/3}}u_\perp+3b_\phi\partial_\perp b_\phi+\frac{3b_\phi^2}{x_\perp}&=&0
\label{Euler2}.
\end{eqnarray}
The combination of the two above equations yields a partial differential equation depending on $u_\perp$ and $b_\phi$:
\begin{eqnarray}
&&u_\perp-\tau^2\partial_\perp\left(\frac{u_\perp}{x_\perp}\right)-\tau^2\partial_\perp^2u_\perp-
\tau\partial_\tau u_\perp+3\tau^2\partial_\tau^2u_\perp\nonumber\\&&
-\frac{3\tau^{7/3}}{x_\perp\epsilon_c}b_\phi^2-\frac{3\tau^{7/3}}{4\epsilon_c}\partial_\perp
b_\phi^2-\frac{9\tau^{10/3}}{4x_\perp\epsilon_c}\partial_\tau b_\phi^2-\frac{3\tau^{10/3}}
{4\epsilon_c}\partial_\perp\partial_\tau b_\phi^2=0.
\label{eq.19}
\end{eqnarray}
For $b_\phi=0$, Eq.~(\ref{eq.19}) is a homogeneous partial differential equation, which can be solved
by separation of variables. The general solution is
\begin{eqnarray}
u_\perp^h(\tau, x_\perp)&=&\sum_k\Big( c_1^k J_1(k x_\perp)+c_2^k Y_1(k x_\perp)\Big)\nonumber\\&&
\times\Big(c_1'^k \tau^{2/3}J_{1/3}(k\tau/\sqrt{3})+c_2'^k \tau^{2/3}Y_{1/3}(k\tau/\sqrt{3})\Big),
\label{eq.20}
\end{eqnarray}
where $k$ can be real or imaginary numbers, $c_{1,2}^k$ and $c_{1,2}'^k$ are integration constants.

For non-vanishing $b_\phi$ we assume a space-time profile of the magnetic field in central collisions in the form:
\begin{eqnarray}
b_\phi^2(\tau, x_\perp)=B_c^2 \tau^{n}\sqrt{\alpha} x_\perp e^{-\alpha x_\perp^2}.
\label{magnetic}
\end{eqnarray}
We see that the magnitude of $b_\phi$ is zero at $x_\perp=0$.
In order to find solutions for transverse velocity $u_\perp$ and energy density $\epsilon$ consistently with the assumed
magnetic field, we found it convenient to first expand the magnetic field, Eq.~(\ref{magnetic})  into a series of
$x_\perp$-dependent functions:
\begin{eqnarray}
b_\phi^2(\tau, x_\perp)=\sum_k \tau^n B_k^2\ f(k x_\perp),
\label{eq.21}
\end{eqnarray}
where $k\geq1$ are now real integers and $B_k^2$ are constants. For simplicity, we have assumed the time
dependence of the magnetic field square as $\tau^n$ with $n<0$, which approximately characterizes the
decay of the magnetic field in heavy ion collisions. This is our key to convert the solution of the
partial differential equation (\ref{eq.19}) into a summation of solutions of ordinary differential equations.

Moreover we replace the solution (\ref{eq.20}) for the radial velocity  $u_\perp(\tau,x_\perp)$, which is valid for $b_\phi=0$,
with the following Ansatz\cite{pu}:
\begin{eqnarray}
u_\perp(\tau, x_\perp)=\sum_k\Big(a_k(\tau)J_1(kx_\perp)+b_k(\tau)Y_1(k x_\perp)\Big).
\label{eq.22}
\end{eqnarray}
It mantains the $x_\perp$ dependence of eq.~(\ref{eq.20}), but embodies the $\tau$ dependence in the coefficients of the Bessel functions.
Note that from the initial condition $u_\perp(\tau, x_\perp=0)=0$ it follows that
$b_k(\tau)=0$.

Now we can substitute the Eqs. (\ref{eq.21}) and (\ref{eq.22}) into Eq. (\ref{eq.19}) and end up with the
equation (at fixed $k$):
\begin{eqnarray}
&&J_1(kx_\perp)\Big(1+\tau^2 k^2-\tau\partial_\tau+3\tau^2\partial_\tau^2\Big)a_k(\tau)\nonumber\\&&
-\frac{3\tau^{7/3+n}}{4\epsilon_c}
 B_k^2 \left(\frac{f(x_\perp)}{x_\perp}(4+3n)+\partial_\perp(f(x_\perp))k(1+n)\right)=0.
 \label{eq.24}
\end{eqnarray}

Here we can apply separation of variables, thus obtaining the following ordinary
differential equation for the function $f(k x_\perp)$:
\begin{eqnarray}
(1+n)k x_\perp\partial_\perp f(k x_\perp)+(4+3n)f(k x_\perp)=k x_\perp J_1(kx_\perp).
\label{eq.25}
\end{eqnarray}
Its general solution is given by
\begin{eqnarray}
f(kx_\perp)&=& \frac{k^2 x_\perp^2 \Gamma \left(\frac{2 n k+2 k+3 n+4}{2 n k+2 k}\right) \,
_1F_2\left(\frac{2 n k+2 k+3 n+4}{2 n k+2 k};2,\frac{4 n k+4 k+3 n+4}{2 n k+2 k};-\frac{1}{4} k^2
x_\perp^2\right)}{4 (n+1)\Gamma \left(\frac{4 k n+4 k+3 n+4}{2 k n+2 k}\right)}\nonumber\\&&
+d_1 (k^2 (n+1) x_\perp)^{-\frac{3 n+4}{k n+k}},
\end{eqnarray}
where $_1F_2$ is the hypergeometric function. The first term is a well-defined function, but the second
one diverges in $x_\perp=0$ for any $n$ except $n=-4/3$ which will be considered in details; hence, $d_1$ must be zero.
For two values of the parameter $n$, $n=-1$ and $n=-4/3$, the solution of Eq.~(\ref{eq.25}) takes a simple form:
\begin{eqnarray}
f(kx_\perp)= k x_\perp J_1(k x_\perp) \,\,\,\,\, (\mathrm{for}~~ n=-1)
\label{eq.27}
\end{eqnarray}
and
\begin{eqnarray}
f(kx_\perp)= d_2+3J_0(k x_\perp) \,\,\,\,\, (\mathrm{for}~~  n=-4/3).
\label{eq.28}
\end{eqnarray}

In order to implement the orthogonal properties of the Bessel functions, for the case $n=-4/3$ we set $d_2=0$ in Eq.~(\ref{eq.28}).
Then we can easily describe the external magnetic field
as a series of Bessel functions, by restricting ourselves to the cases $n=-1$ and $n=-4/3$.

We write the solution for $n=-1$ as
\begin{eqnarray}
b_\phi^2(\tau, x_\perp)=\sum_{k} \ \tau^{-1}\ B_k^2 \ \beta_{1 k} \frac{x_\perp}{a} \ J_1(\beta_{1 k}
 \frac{x_\perp}{a})
 \label{eq.29}
\end{eqnarray}
where the coefficients $B_k^2$ are given by
\begin{eqnarray}
B_k^2=\frac{2 a}{a^2 \beta_{1 k} [J_2(\beta_{1 k})]^2}\int_0^a \ J_1(\beta_{1 k}\frac{x_\perp}{a})
\ b_\phi^2 \ dx_\perp,
\label{eq.30}
\end{eqnarray}
 $\beta_{1 k}$ being the $k$th zero of $J_1$.

 For $n=-4/3$ the solution for the magnetic field can be written as
\begin{eqnarray}
b_\phi^2(\tau, x_\perp)=\sum_{k} \ \tau^{-4/3}\ B_k^2 \ 3 \ J_0\left(\beta_{0 k} \frac{x_\perp}{a}\right)
\label{eq.31}
\end{eqnarray}
where the coefficients $B_k^2$ are given by
\begin{eqnarray}
B_k^2=\frac{2}{3 \ a^2[J_1(\beta_{0 k})]^2}\int_0^a \ x_\perp J_0(\beta_{0 k}\frac{x_\perp}{a})
\ b_\phi^2 \ dx_\perp
\label{eq.32}
\end{eqnarray}
 $\beta_{0 k}$ being the $k$th zero of $J_0$; in the above $k=\beta_{ik}/a$ ($i=0,1$).

Finally the coefficients $a_k(\tau)$ in Eq.~(\ref{eq.22}) can be obtained by solving the
following ordinary differential equation:
\begin{eqnarray}
\left(k^2 \tau^2+1\right) a_k(\tau)+\tau \left(3 \tau a_k''(\tau)-a_k'(\tau)\right)-
\frac{3 k B_k^2 \tau^{n+\frac{7}{3}}}{4 \epsilon _c}=0.
\end{eqnarray}

The analytical solution for $n=-1$ is
\begin{eqnarray}
a_k(\tau)&=&c_1^k \tau ^{2/3} J_{\frac{1}{3}}(\frac{k \tau }{\sqrt{3}})+c_2^k
\tau ^{2/3} Y_{\frac{1}{3}}(\frac{k \tau }{\sqrt{3}})
+\frac{\pi k B_k^2}{48 \Gamma (\frac{2}{3}) \Gamma (\frac{7}{6}) \Gamma (\frac{4}{3})
\epsilon _c \sqrt[3]{k \tau }} \nonumber\\&&
\Big(-2^{2/3} \sqrt[3]{3} \tau ^{4/3} \Gamma (\frac{2}{3}) \Gamma (\frac{7}{6}) (k
\tau )^{2/3} J_{\frac{1}{3}}(\frac{k \tau }{\sqrt{3}})\, _1F_2(\frac{1}{2};\frac{4}{3},
\frac{3}{2};-\frac{1}{12} k^2 \tau ^2)\nonumber\\&&
+2 \sqrt[3]{2} 3^{2/3} \tau ^{4/3} \Gamma(\frac{4}{3}) \Gamma (\frac{1}{6}) J_{\frac{1}{3}}
(\frac{k \tau }{\sqrt{3}}) \, _1F_2(\frac{1}{6};\frac{2}{3},\frac{7}{6};-\frac{1}{12} k^2 \tau ^2)
\nonumber\\&&
+2^{2/3} 3^{5/6} \tau ^{4/3} \Gamma (\frac{2}{3}) \Gamma (\frac{7}{6}) (k \tau )^{2/3}
Y_{\frac{1}{3}}(\frac{k \tau }{\sqrt{3}})\, _1F_2(\frac{1}{2};\frac{4}{3},\frac{3}{2};-
\frac{1}{12} k^2 \tau ^2)\Big)
\end{eqnarray}
while for $n=-4/3$ the solution is
\begin{eqnarray}
a_k(\tau)&=&c_1^k \tau ^{2/3} J_{\frac{1}{3}}(\frac{k \tau }{\sqrt{3}})+c_2^k \tau ^{2/3}
Y_{\frac{1}{3}}(\frac{k \tau }{\sqrt{3}})
+\frac{\pi  k \tau  B_k^2}{96 \Gamma^2
 (\frac{4}{3})\epsilon _c \sqrt[3]{k \tau }}\nonumber\\&&
 \Big(-2^{2/3} \sqrt[3]{3} \Gamma(\frac{1}{3})(k \tau )^{2/3} J_{\frac{1}{3}}
(\frac{k \tau }{\sqrt{3}})\, _1F_2(\frac{1}{3};\frac{4}{3},\frac{4}{3};-\frac{1}{12} k^2
\tau ^2)\nonumber\\&&
 +2^{2/3} 3^{5/6} \Gamma(\frac{1}{3})(k\tau )^{2/3} Y_{\frac{1}{3}}(\frac{k \tau }{\sqrt{3}})
\,_1F_2(\frac{1}{3};\frac{4}{3},\frac{4}{3};-\frac{1}{12} k^2 \tau ^2)\nonumber\\&&
 -4 \sqrt[3]{2} 3^{2/3} \Gamma^2(\frac{4}{3}) J_{\frac{1}{3}}(\frac{k\tau }{\sqrt{3}})
G_{1,3}^{2,0}(\frac{k^2 \tau ^2}{12}|
\begin{array}{c}
 1 \\
 0,0,\frac{1}{3} \\
\end{array}
)\Big).
\end{eqnarray}
In the above $G_{mn}^{pq}$ is the Meijer function.

The transverse velocity then takes the form
\begin{eqnarray}
u_\perp(\tau, x_\perp)=\sum_k a_k(\tau) J_1( k x_\perp).
\end{eqnarray}
In order to completely determine the function $u_\perp(\tau, x_\perp)$ we must fix the
integration constants $c_1^k$ and $c_2^k$. It is convenient to consider the boundary conditions
at $\tau\rightarrow\infty$. Since
$b_\phi^2(\infty, x_\perp)\rightarrow0$ we expect $u_\perp(\infty, x_\perp)\rightarrow0$.
By making late-time expansion of $u_\perp$, one finds that $u_\perp$ takes the asymptotic
form $f(\tau)\tau^{1/6}$ where $f(\tau)$ is an oscillatory function. In order to prevent
divergencies of the transverse velocity one has to impose that the coefficient
of $\tau^{1/6}$ is equal to zero.
The solutions satisfying these boundary condition at $\tau\rightarrow\infty$ are shown in the following.

For $n=-1$,
\begin{eqnarray}
 c_1^k = \frac{\sqrt[3]{k} (3 \pi ^{3/2} \Gamma (\frac{7}{6})-\sqrt{\pi } \Gamma^2(\frac{1}{6})
\Gamma(\frac{5}{6})) B_k^2}{24 \sqrt[3]{2} \sqrt[6]{3} \Gamma (\frac{5}{6}) \Gamma (\frac{7}{6}) \epsilon _c}, \,
 c_2^k = -\frac{\sqrt[3]{\frac{3}{2}}\pi ^{3/2}\sqrt[3]{k} B_k^2}{8 \Gamma(\frac{5}{6})\epsilon _c}.
\end{eqnarray}
For $n=-4/3$,
\begin{eqnarray}
c_1^k = \frac{\pi k^{2/3} \Gamma(\frac{1}{3})^2 B_k^2}{24\ 2^{2/3} \sqrt[3]{3} \Gamma(\frac{4}{3}) \epsilon _c}, \,
 c_2^k = -\frac{\pi  k^{2/3} \Gamma(\frac{1}{3})^2 B_k^2}{8\ 2^{2/3} 3^{5/6} \Gamma(\frac{4}{3}) \epsilon _c}.
\end{eqnarray}
After obtaining $u_\perp(\tau, x_\perp)$ we can get, correspondingly, the modified energy density
from Eq.~(\ref{Euler2}).
For $n=-1$, it reads:
\begin{eqnarray}
&&\epsilon_1(\tau,x_\perp)= \sum_k h_k(\tau)+\sum_k  \frac{1}{24 k \tau^{7/3}}\Big(32
\epsilon _c [J_0(k x_\perp)-1] [a_k(\tau)-3 t a_k'(\tau)]\nonumber\\&&~
-9 B_k^2 k \tau^{4/3} \left[k^2 x_\perp^2 \, _0F_1(2;-\frac{1}{4} k^2 x_\perp^2)+2 k x_\perp J_1
(k x_\perp)-8 J_0(k x_\perp)+8\right]\Big),\label{eq.39}
\end{eqnarray}
where $h(\tau)$ is the constant of integration and can be obtained form Eq.~(\ref{Euler1}). We find,
\begin{eqnarray}
h_k(\tau)=\frac{\int_1^\tau \frac{4}{3} k \epsilon _c a_k(s) \, ds}{\tau^{4/3}}.
\label{eq.40}
\end{eqnarray}
For $n=-4/3$, instead,
\begin{eqnarray}
\epsilon_1 (\tau,x_\perp)&=& \sum_k h_k(\tau)+\sum_k \frac{1}{6 k \tau^{7/3}}\Big(8
\epsilon _c[J_0(k x_\perp)-1] (a_k(\tau)-3 t a_k'(\tau))\nonumber\\&&
-27 B_k^2 k \tau [-G_{1,3}^{2,0}(\frac{k^2 x_\perp^2}{4}|
\begin{array}{c}
 1 \\
 0,0,0 \\
\end{array}
)+J_0(k x_\perp)-1]\Big),
\label{eq.41}
\end{eqnarray}
where
\begin{eqnarray}
h_k(\tau)=\frac{\int_1^\tau \left(\frac{4}{3} k \epsilon _c a_k(s)-\frac{B_k^2}{s}\right)
\, ds}{\tau^{4/3}}.
\label{eq.42}
\end{eqnarray}
Note that, the integrals (\ref{eq.40}) and (\ref{eq.42}) should be calculated numerically.
%*********************************************************************************
\section{Results and discussion}

In this Section we will present the transverse velocity and energy density numerically obtained from
our perturbation approach: this two quantities will help in understanding  the space time evolution
of the quark-gluon plasma in heavy ion collisions.
The typical magnetic field produced in Au-Au  peripheral collisions at $\sqrt{s_{NN}}=200$~GeV reaches
$ |e B| \sim10 m_{\pi}^{2}$.  The estimate $\epsilon\sim5.4~ $GeV/fm$^{3}$ at a proper time of about
$\tau=1$~fm is taken from \cite{gubser}. By taking $m_\pi\approx150$ MeV and $e^2=4 \pi /137$,
one finds $B^2/\epsilon_c\sim0.6$. This value in central collisions is much smaller than in peripheral
collisions; therefore, in our calculations we assumed the even smaller value $B_c^2/\epsilon_c=0.1$,
which correspond to $\sigma\sim0.015$. Note that in our calculations any change in the ratio $B_c^2/\epsilon_c$
will only scale the solutions.
We will use cylindrical coordinates whose longitudinal component is chosen to be the third component of
Cartesian coordinate, e.g., $\vec{x}=(x_\perp,\phi,z)$.

\subsection{Numerical solution for the case $n=-1$}
\begin{figure}[t!]
\begin{center}
\includegraphics[width=4in]{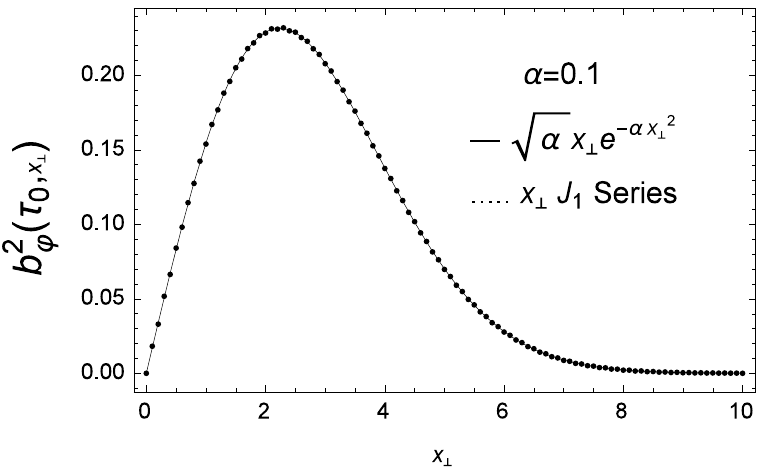}\\
\end{center}
\caption{A comparison between the approximated $b^2_\phi$ in Bessel series (dotted curve) and the
assumed external magnetic field (solid curve) with n=-1.}
\label{fig:fig1}
\end{figure}

The external magnetic field profile Eq.~(\ref{magnetic}) can be reproduced by expressing $b_\phi^2$ via a series of
Bessel functions as shown in Eq.~(\ref{eq.29}).
The first ten coefficients of series and on the
$B_k^2$ calculated according to  Eq.~(\ref{eq.30}) for $\alpha=0.1$ are: $B_c^2$\{0.112499, \,0.111212,\,
0.0707575, \,0.0391739, \,0.0231679, \,0.0153799, \,0.0110821, \, 0.0084056, \,0.00661182, \,0.00534074\}.
In order to reproduce the assumed external magnetic profile Eq.~(\ref{magnetic}) we had to take in the calculation
the first 100 terms of the series.
Fig. \ref{fig:fig1} shows a comparison between the approximated magnetic
field in Bessel series and the assumed magnetic profile Eq.~(\ref{magnetic}). Note that the Fourier
expansion matches the assumed magnetic profile in the whole region of $x_\perp$, hence the solutions
for the radial velocity and the energy density are valid in the entire region $x_\perp\in(0,\infty)$.
\begin{figure}[th!]
\begin{center}
\includegraphics[width=4in]{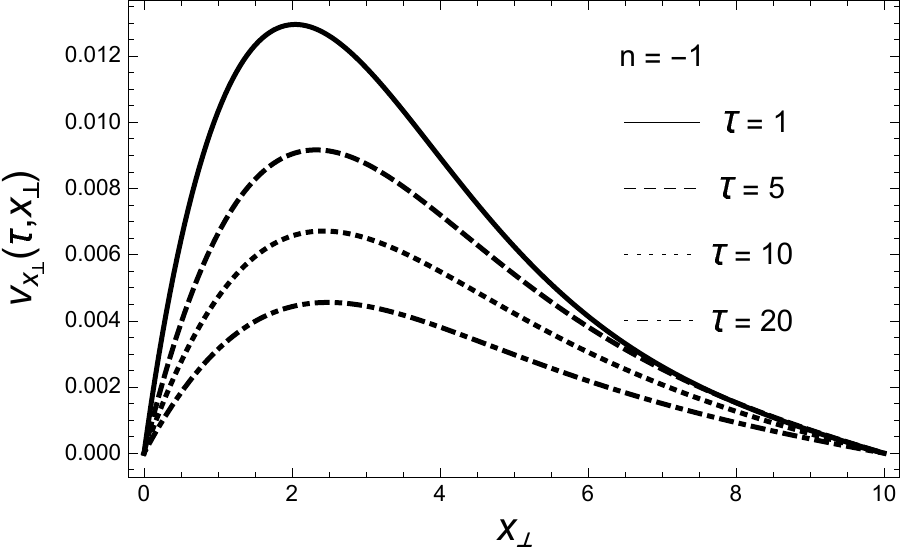}\\
\end{center}
\caption{$v_{x_\perp}$ as a function of $x_\perp$ for diffIn this sense, without loss of generality, we can modify the text accordingly.erent values of $\tau$.}
\label{fig:fig2}
\end{figure}

\begin{figure}[th!]
\begin{center}
\includegraphics[width=4in]{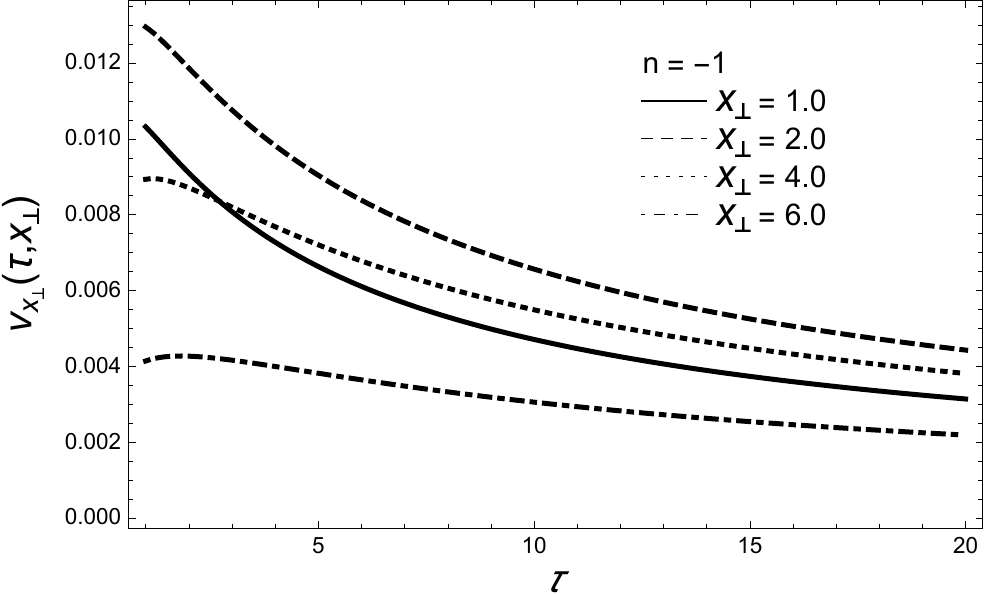}\\
\end{center}
\caption{$v_{x_\perp}$ as a function of $\tau$ for different values of $x_\perp$.}
\label{fig:fig3}
\end{figure}

Next we show plots of the fluid velocity ($v_{x_\perp}\equiv u_\perp/ u_\tau$) and of the energy density
modified by the magnetic field with
$B^2_c/\epsilon_c=0.1$. In Figs.\ref{fig:fig2} and \ref{fig:fig3}  $v_{x_\perp}(\tau,x_\perp)$  is
displayed, at either fixed $\tau$ or fixed $x_\perp$, respectively. From Fig.\ref{fig:fig2}, one finds
that $v_{x_\perp}(\tau,0)=0$ and the radial velocity $v_{x_\perp}$ first increases from $x_\perp=0$,
has a maximum at intermediate $x_\perp$ and then gradually decreases with $x_\perp$. As shown in
Fig.\ref{fig:fig3}, $v_{x_\perp}$ at fixed $x_\perp$ becomes smaller  at late times, due to the decay
of the magnetic field, in agreement with the curves displayed in Fig.\ref{fig:fig2}.
\begin{figure}[th!]
\begin{center}
\includegraphics[width=4in]{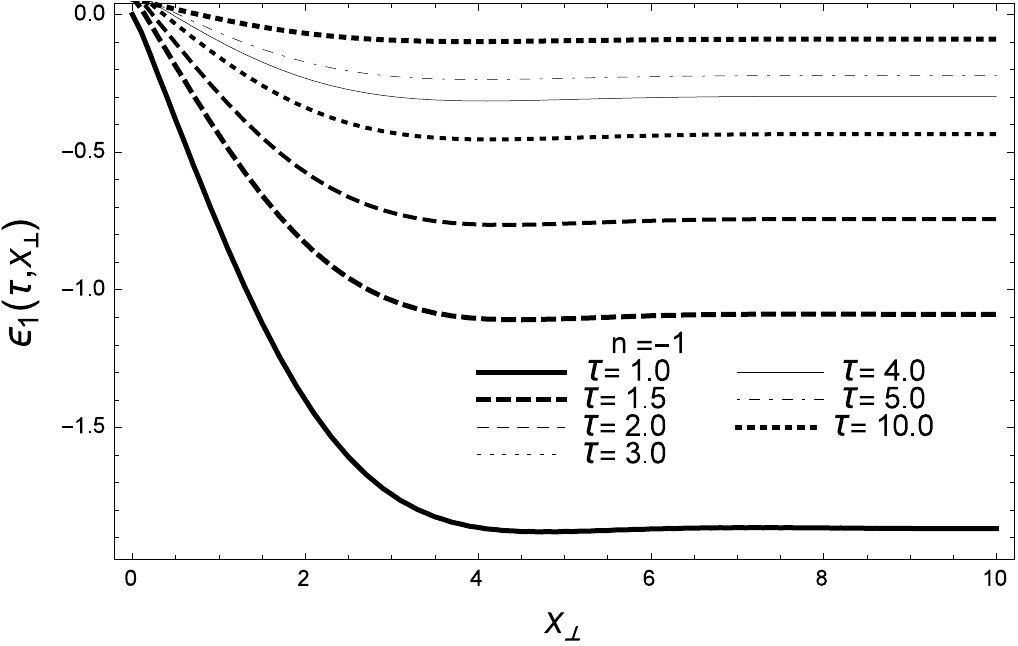} \\
\end{center}
\caption{$\epsilon_1$ as a function of $x_\perp$ for different values of $\tau$.}
\label{fig:fig4}
\end{figure}
\begin{figure}[th!]
\begin{center}
\includegraphics[width=4in]{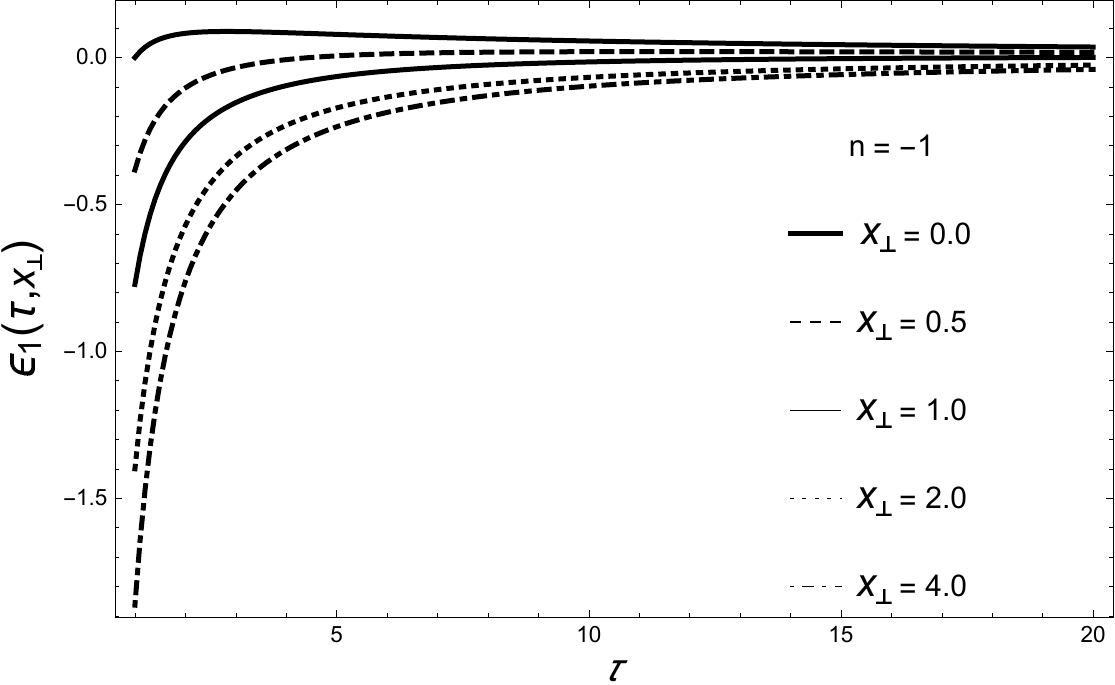} \\
\end{center}
\caption{$\epsilon_1$ as a function of $\tau$ for different values of $x_\perp$.}
\label{fig:fig5}
\end{figure}

Fig.\ref{fig:fig4} shows the correction energy density $\epsilon_1(\tau,x_\perp)$ as a function
of $x_\perp$ for different values of $\tau$; we remind the reader that the total energy density is
$\epsilon= \epsilon_0(\tau)+\epsilon_1(\tau,x_\perp)$ and the latter is the component which is
truly affected by the magnetic field.
Fig. \ref{fig:fig5} shows the correction energy density as a function of $\tau$ for different values
of $x_\perp$. Here we find that for $x_\perp=0$ the correction energy density is positive, starting from
zero at proper time $\tau=1$~fm and showing a shallow maximum; for $x_\perp=0.5$~fm
the correction energy density is negative at $\tau=1$~fm and increases with $\tau$
reaching zero at approximately $\tau=4.5$~fm, becoming then slightly positive. For the other values of $x_\perp$
the correction energy density is negative at any time and monotonically increases toward zero.

\begin{figure}[th!]
\begin{center}
\includegraphics[width=4in]{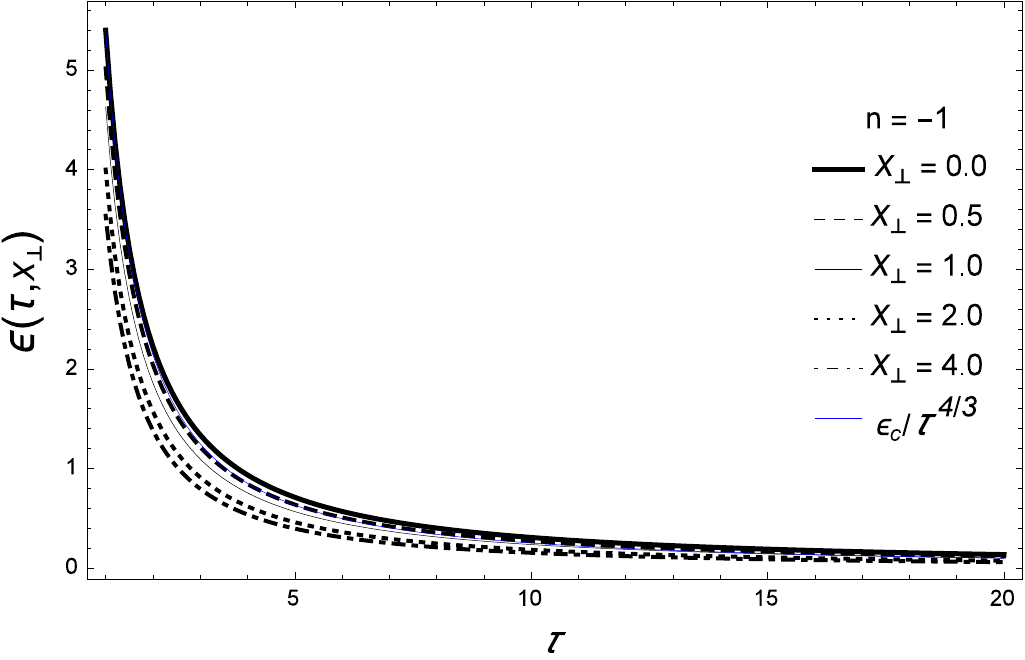} \\
\end{center}
\caption{$\epsilon(\tau,x_\perp)$ as a function of $\tau$ for several values of $x_\perp$.}
\label{fig:fig6}
\end{figure}

This behavior can also be seen in Figs.\ref{fig:fig6} and \ref{fig:fig7} which
show the normal and Log-Log plots of the total energy density $\epsilon(\tau,x_\perp)$ as a function of
$\tau$ for several values of $x_\perp$, respectively. The time evolution of the energy density for different
values of $x_\perp$ in the work of Gubser \cite{gubser} has nearly the same behavior as in
Fig.\ref{fig:fig7}, stemming from a similar trend
of the correction energy density as a function of $\tau$, like the one illustrated in our Fig.\ref{fig:fig5}.
In the Gubser work for $\tau<4.6$~fm the energy density is
positive for $x_\perp\leq3 $~fm and negative for $x_\perp\geq4$~fm and it is negative for any
$x_\perp$ for $\tau>4.6$~fm.
\begin{figure}[th!]
\begin{center}
\includegraphics[width=4in]{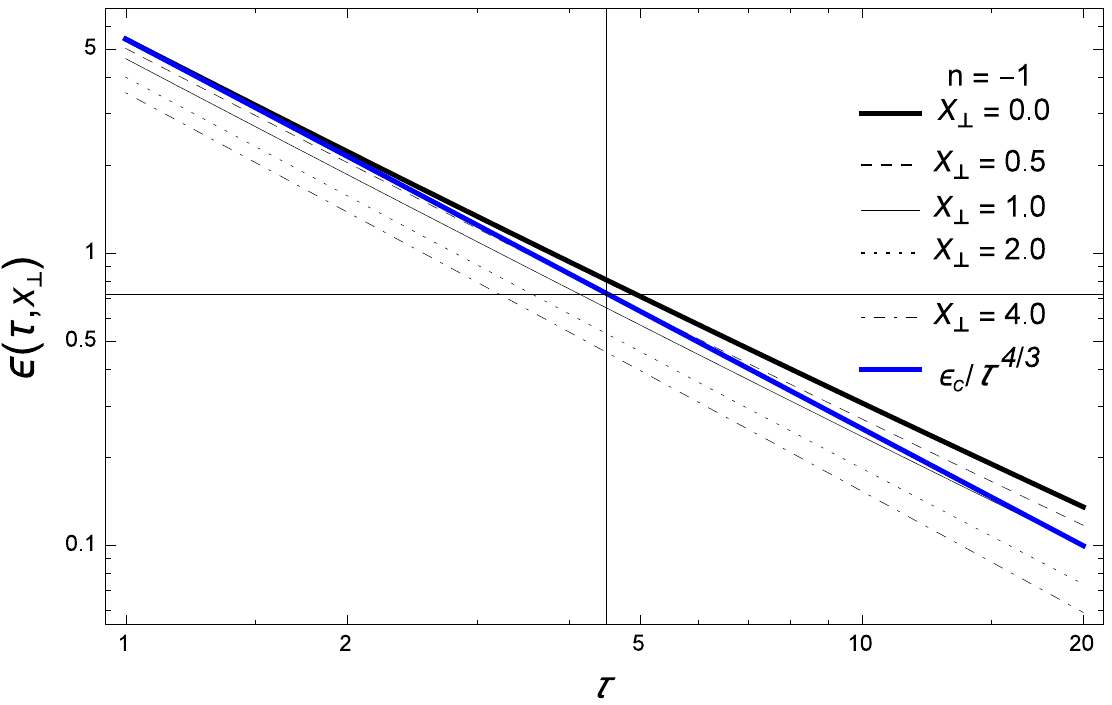} \\
\label{fig:fig7}
\end{center}
\caption{Log-Log plot of $\epsilon(\tau,x_\perp)$ as a function of $\tau$ for several values of $x_\perp$.
The bold blue line shows the dependence $\epsilon/\tau^{4/3}$, where $\epsilon$ is in GeV/fm$^3$ and
$\tau$ in fm. We have chosen $\epsilon=5.4$ GeV/fm$^3$ at $\tau=1$~fm from \cite{gubser}.}
\label{fig:fig7}
\end{figure}
\begin{figure}[th!]
\begin{center}
\includegraphics[width=4in]{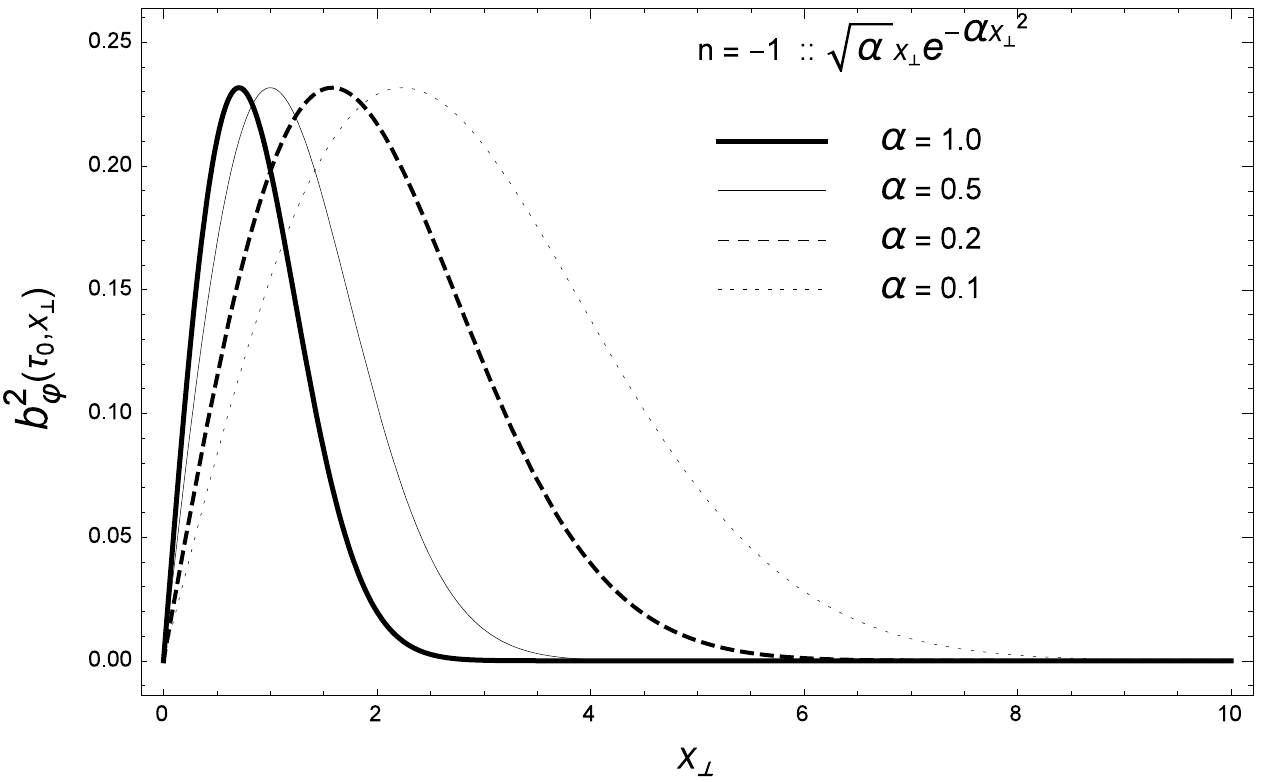} \\
\end{center}
\caption{$b^2_\phi(\tau,x_\perp)$ as a function of $x_\perp$ for different values of $\alpha$ at $\tau_0=1$~fm.}
\label{fig:figcom1}
\end{figure}\\
\begin{figure}[th!]
\begin{center}
\includegraphics[width=4in]{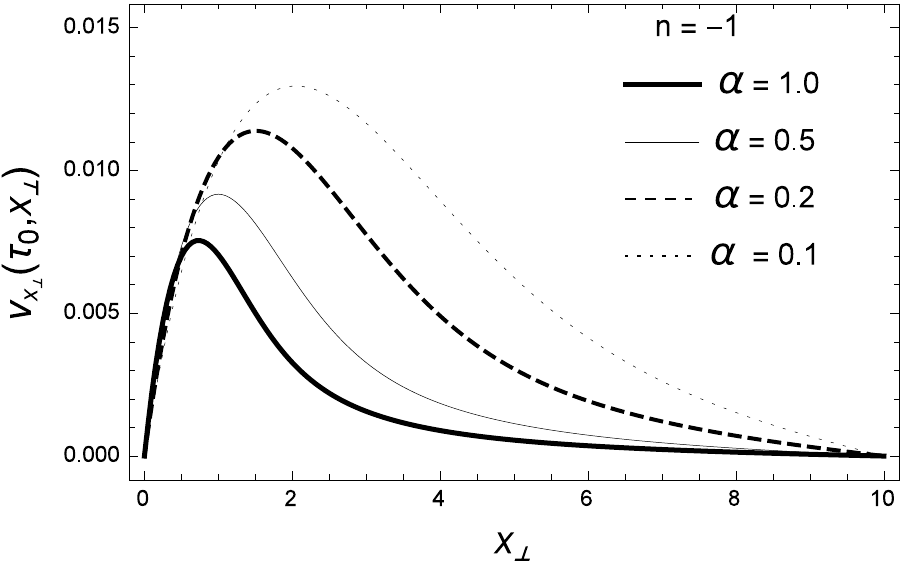} \\
\end{center}
\caption{$v_{x_\perp}(\tau,x_\perp)$ as a function of $x_\perp$ for different values of $\alpha$ at $\tau_0=1$~fm.}
\label{fig:figcom2}
\end{figure}\\
\begin{figure}[th!]
\begin{center}
\includegraphics[width=4in]{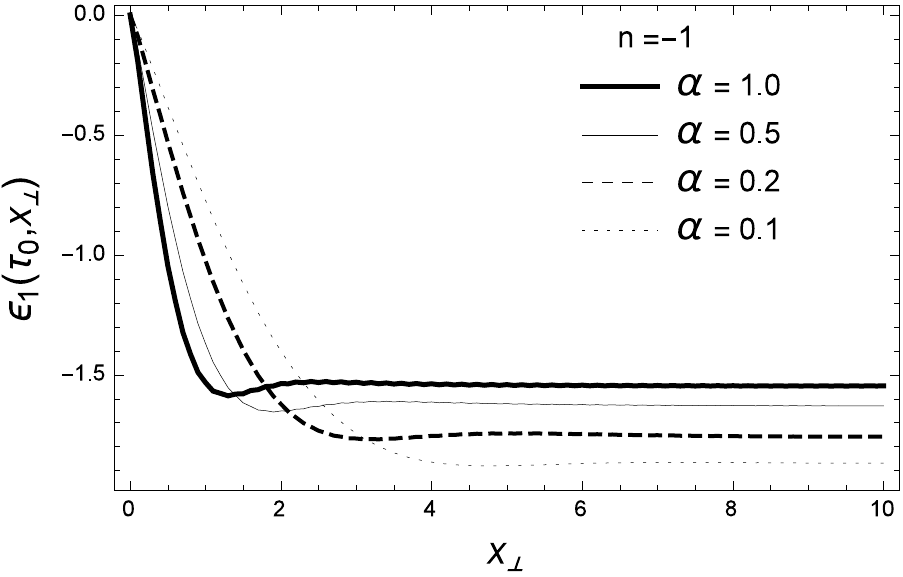} \\
\end{center}
\caption{$\epsilon_1(\tau,x_\perp)$ as a function of $x_\perp$ for different values of $\alpha$ at $\tau_0=1$~fm.}
\label{fig:figcom3}
\end{figure}

It is interesting to investigate variations of the spatial width of the external magnetic field: this affects the
Fourier series which reproduces the assumed distribution for the magnetic field; moreover we find that
$v_{x_\perp}(\tau,x_\perp)$ and $\epsilon_1(\tau,x_\perp)$ have an important dependence on the parameter
$\alpha$ (with dimension square of inverse length), which characterizes the spatial width of the magnetic field.
In Fig. \ref{fig:figcom1}, we plot the external
magnetic profile at $\tau=1$~fm for several different values of $\alpha$. In Figs. \ref{fig:figcom2}
and \ref{fig:figcom3}, we plot $v_{x_\perp}$ and $\epsilon_1$  at $\tau=1$~fm for references.
The $v_{x_\perp}$ gets smaller when $\alpha$ is increased. It seems that the parameter $\alpha$ plays the
role of the parameter $1/q^2$ in Ref.\cite{gubser}:indeed the radial flow velocity (versus
$x_\perp$ at $\tau=0.6$~fm)  becomes smaller when $1/q$ increases.
\begin{figure}[th!]
\begin{center}
\includegraphics[width=4in]{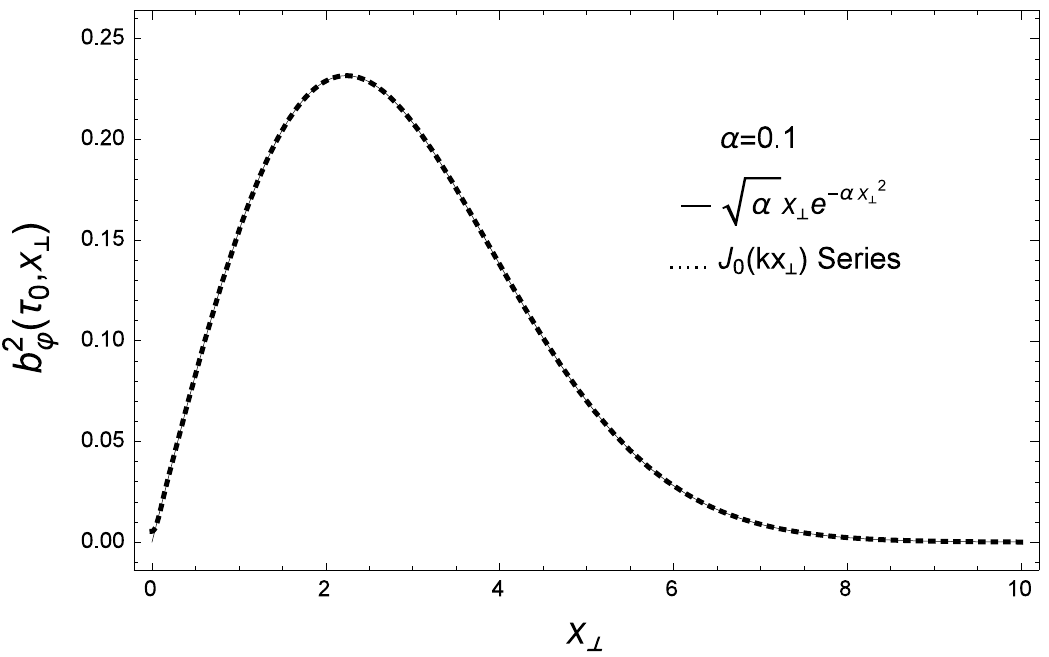}\\
\end{center}
\caption{A comparison between the approximated $b^2_\phi$ in Bessel series (dotted curve) and the
assumed external magnetic field (solid curve) with n=-4/3.}
\label{fig:fig11}
\end{figure}
\begin{figure}[th!]
\begin{center}
\includegraphics[width=4in]{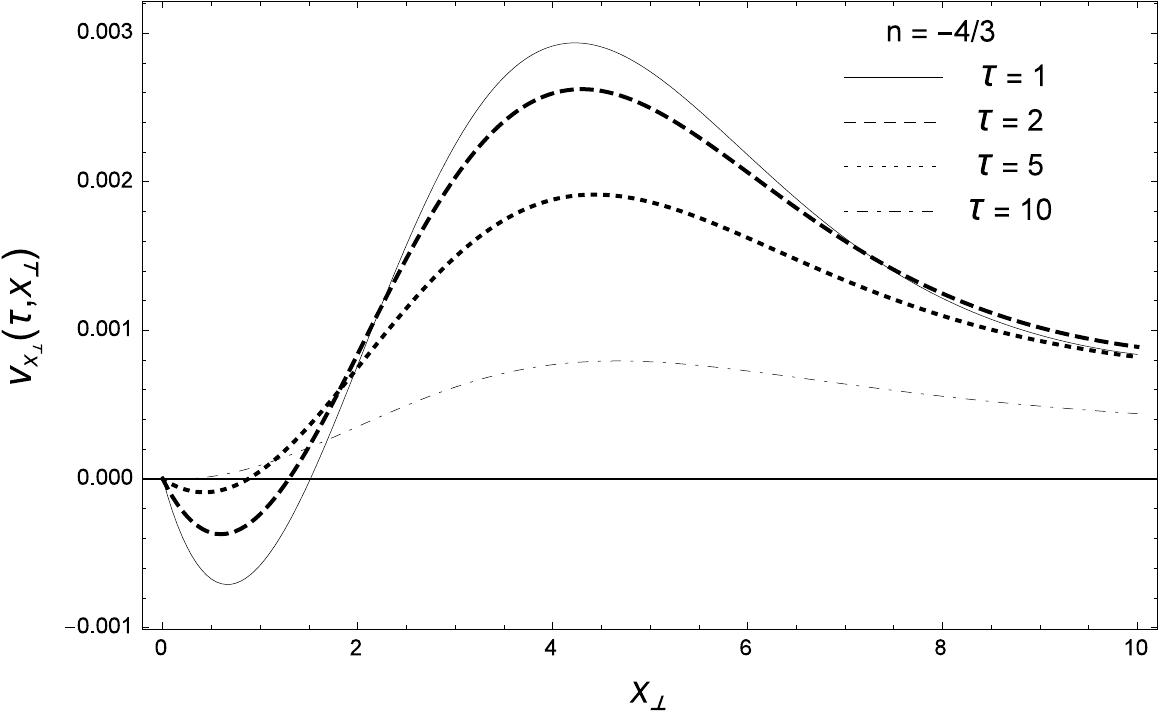} \\
\end{center}
\caption{$v_{x_\perp}$ as a function of $x_\perp$ for different values of $\tau$.}
\label{fig:fig22}
\end{figure}
\begin{figure}[th!]
\begin{center}
\includegraphics[width=4in]{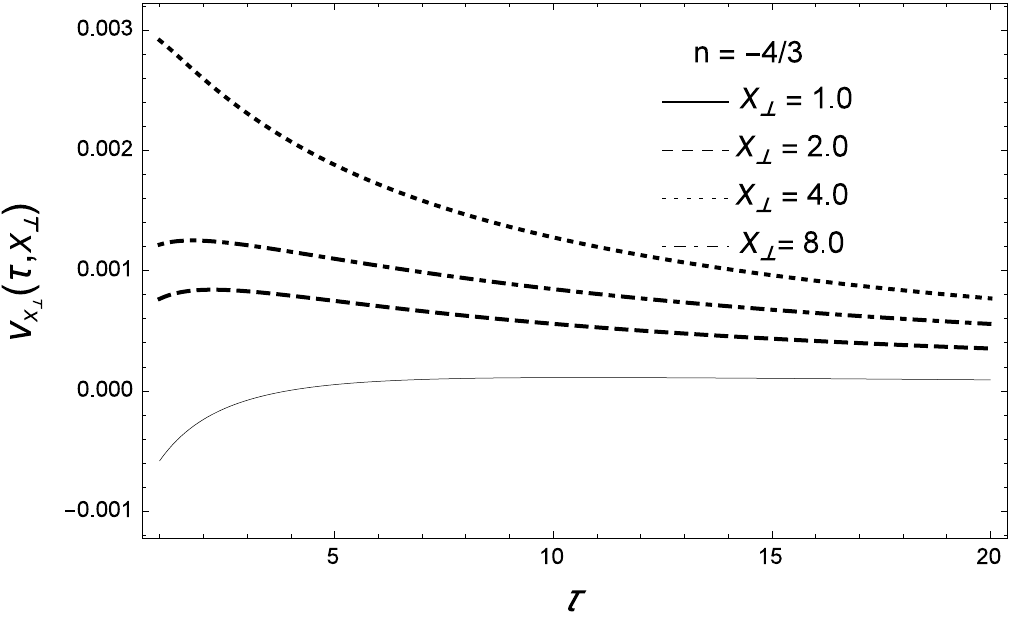}\\
\end{center}
\caption{$v_{x_\perp}$ as a function of $\tau$ for different values of $x_\perp$.}
\label{fig:fig33}
\end{figure}
\begin{figure}[th!]
\begin{center}
\includegraphics[width=4in]{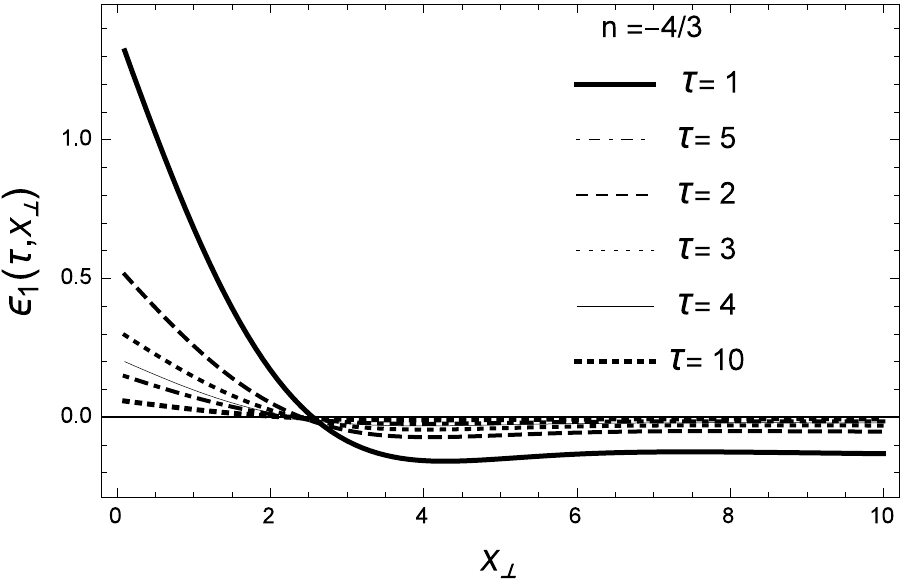}\\
\end{center}
\caption{$\epsilon_1$ as a function of $x_\perp$ for different values of $\tau$.}
\label{fig:fig44}
\end{figure}

\subsection{Numerical solution for the case $n=-4/3$}
For the case $n=-4/3$, the external magnetic field profile Eq.~(\ref{magnetic}) can be reproduced as a series of Bessel functions as shown in
Eq.~(\ref{eq.31}). The first 10 coefficients
$B_k^2$ calculated according to of Eq.~(\ref{eq.32}) for $\alpha=0.1$ are: $B_c^2$\{0.04745,\,0.0371507,\,-0.00832931,\,-0.0215885,\,-0.0146208,\,-0.00825935,\,-0.00515007,\,-0.00362071,\,-0.00269932,\,-0.00212629\}.
In order to reproduce the assumed external magnetic profile Eq.~(\ref{magnetic}), one may take the first 100 terms of the series in the calculation.
Fig. \ref{fig:fig11} shows a comparison between the approximated magnetic field by the Bessel series
and the assumed magnetic profile Eq.~(\ref{magnetic}).\\
Figs. \ref{fig:fig22} and \ref{fig:fig33} show $v_{x_\perp}(\tau,x_\perp)$ at either fixed
$\tau$ or fixed $x_\perp$, respectively. The qualitative behaviors of $v_{x_\perp}(\tau,x_\perp)$
in both figures are different from the case $n=-1$ and the amplitude is smaller.
While for $n=-1$, the direction of the fluid velocity is always positive, for $n=-4/3$ the direction of fluid
velocity changes during the expansion of the fluid.
\begin{figure}[th!]
\begin{center}
\includegraphics[width=4in]{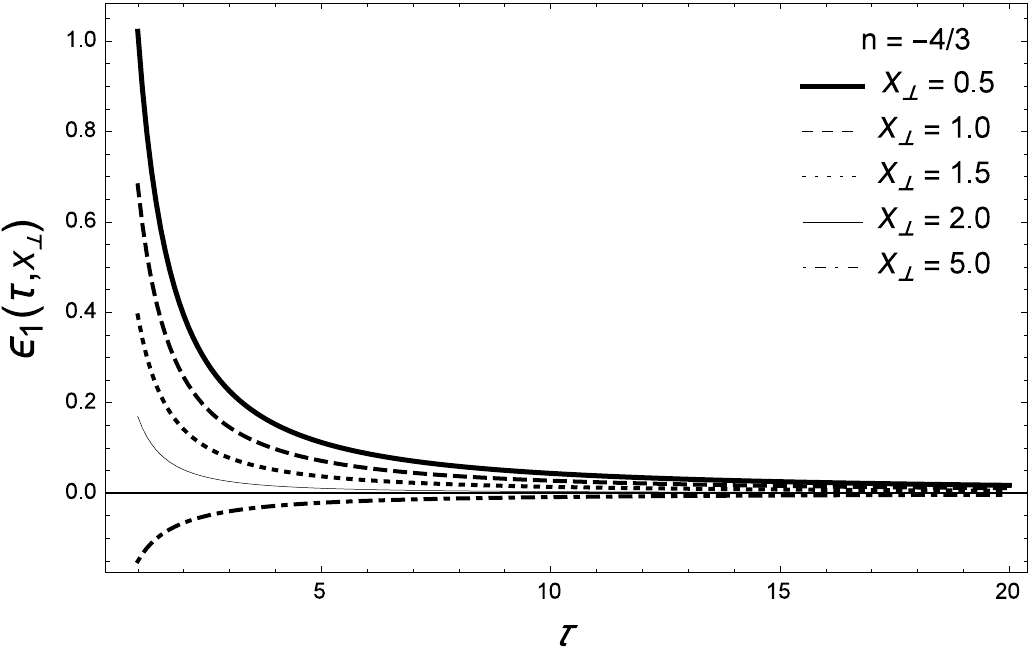}\\
\end{center}
\caption{$\epsilon_1$ as a function of $\tau$ for different values of $x_\perp$.}
\label{fig:fig55}
\end{figure}

\begin{figure}[th!]
\begin{center}
\includegraphics[width=4in]{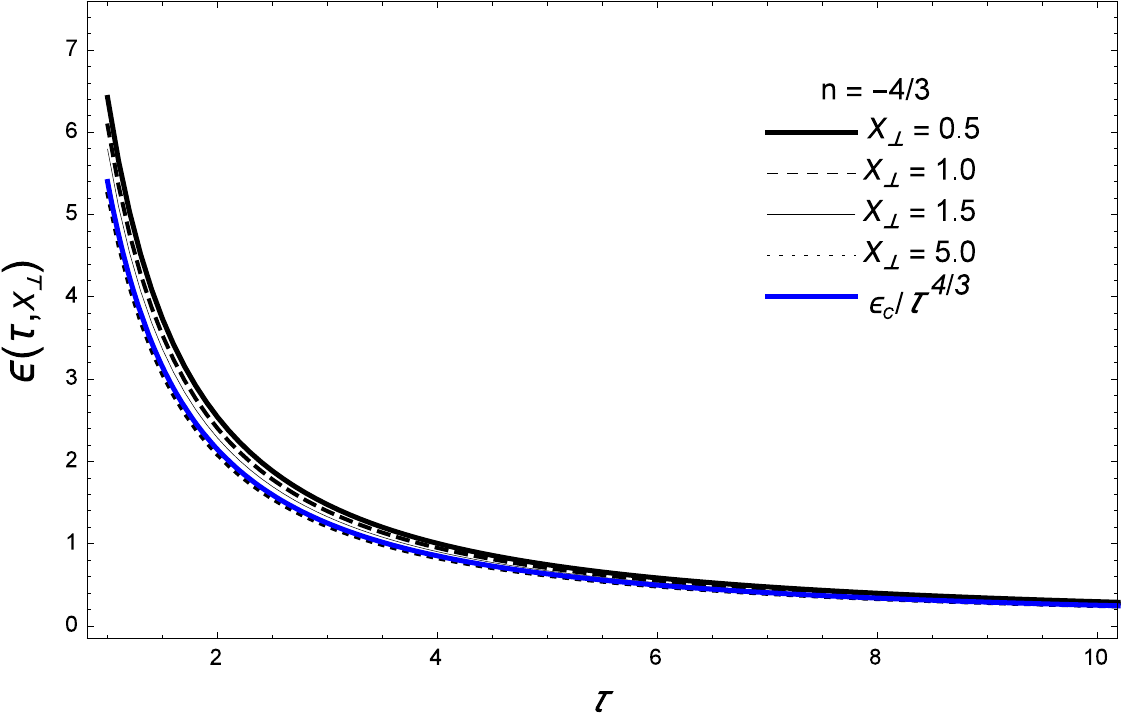} \\
\end{center}
\caption{$\epsilon(\tau,x_\perp)$ as a function of $\tau$ for several values of $x_\perp$.}
\label{fig:fig66}
\end{figure}

\begin{figure}[th!]
\begin{center}
\includegraphics[width=4in]{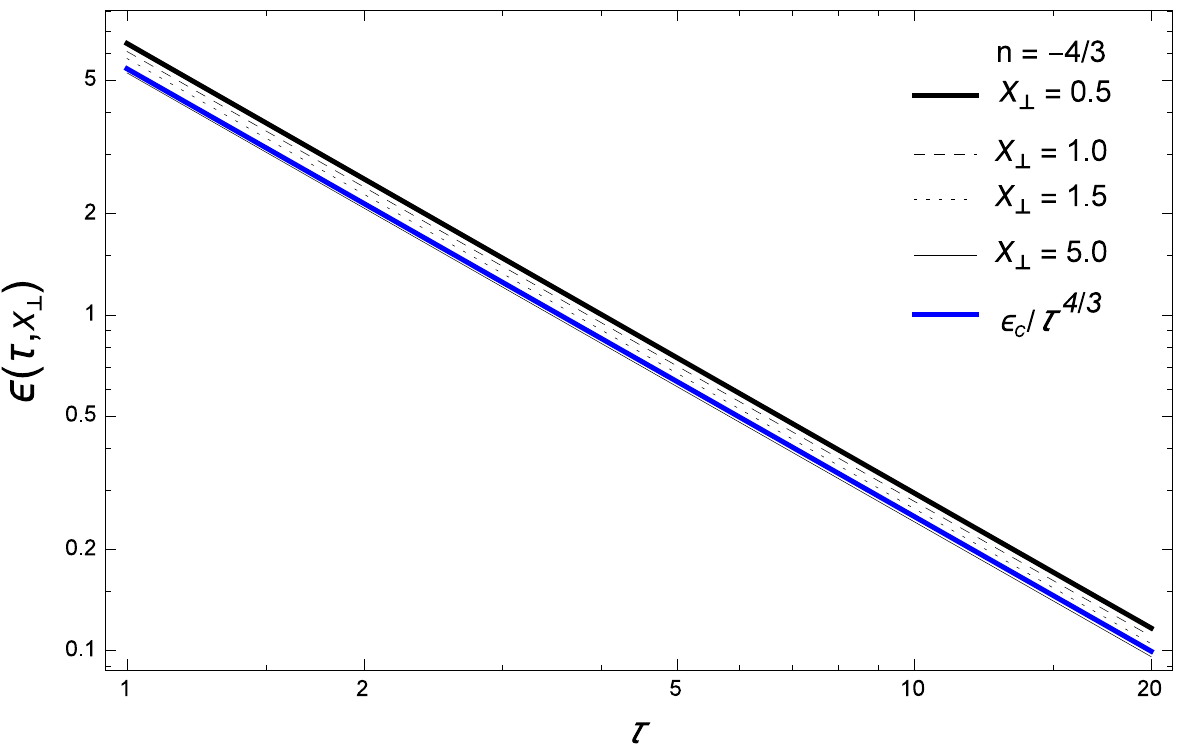} \\
\end{center}
\caption{Log-Log plot of $\epsilon(\tau,x_\perp)$ as a function of $\tau$ for several values of $x_\perp$.
The bold blue line shows the dependence $\epsilon/\tau^{4/3}$, where $\epsilon$ is in GeV/fm$^3$ and $\tau$
in fm. We have chosen $\epsilon=5.4 $ GeV/fm$^3$ at $\tau=1$~fm from \cite{gubser}.}
\label{fig:fig77}
\end{figure}

Fig.~\ref{fig:fig44} shows the correction energy density as a function of $x_\perp$ for
different values of $\tau$. Fig.~\ref{fig:fig55} shows the correction energy density as
a function of $\tau$ for different values of $x_\perp$. Figs. \ref{fig:fig66} and \ref{fig:fig77}
show the normal and Log-Log plots of the total $\epsilon(\tau,x_\perp)$ as a function of $\tau$ for several
values of $x_\perp$, respectively. In Fig. \ref{fig:fig44}, we find that for $x_\perp=0$ the correction
energy density is always positive and then it decreases from the value at $x_\perp=0$ with increasing $x_\perp$.
From Fig. \ref{fig:fig55} one also finds that for the case $n=-4/3$
the correction energy density is always positive, at variance with the case $n=-1$. The same feature can be
obviously extracted from Figs. \ref{fig:fig66} and \ref{fig:fig77}.

\begin{figure}[th!]
\begin{center}
\includegraphics[width=4in]{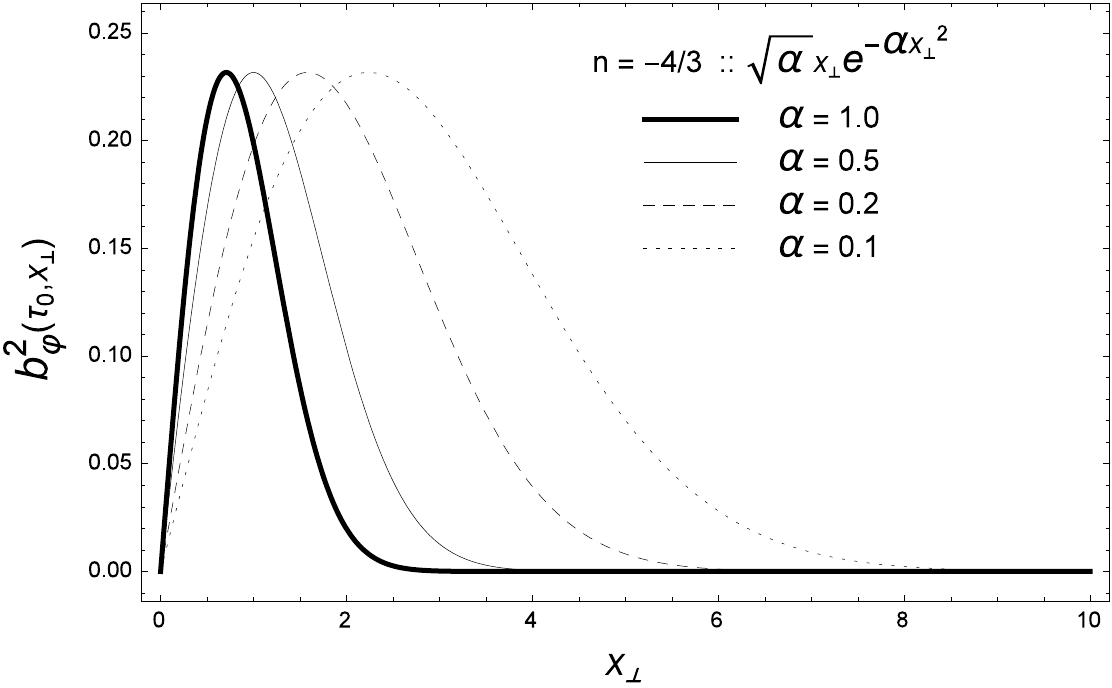} \\
\end{center}
\caption{$b^2_\phi(\tau,x_\perp)$ as a function of $x_\perp$ for different values of $\alpha$ at $\tau_0=1$~fm.}
\label{fig:figcom11}
\end{figure}

\begin{figure}[th!]
\begin{center}
\includegraphics[width=4in]{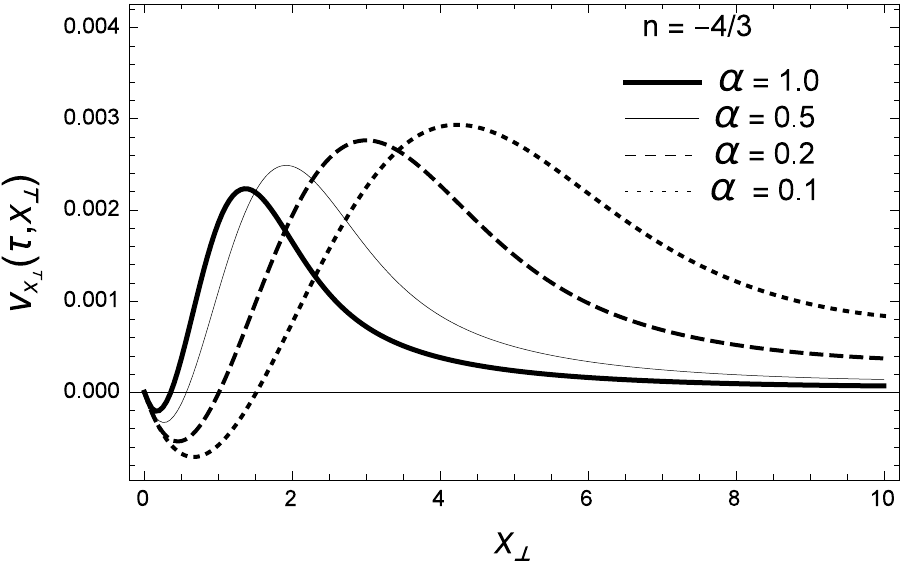} \\
\end{center}
\caption{$v_{x_\perp}(\tau,x_\perp)$ as a function of $x_\perp$ for different values of $\alpha$ at $\tau_0=1$~fm.}
\label{fig:figcom22}
\end{figure}

\begin{figure}[th!]
\begin{center}
\includegraphics[width=4in]{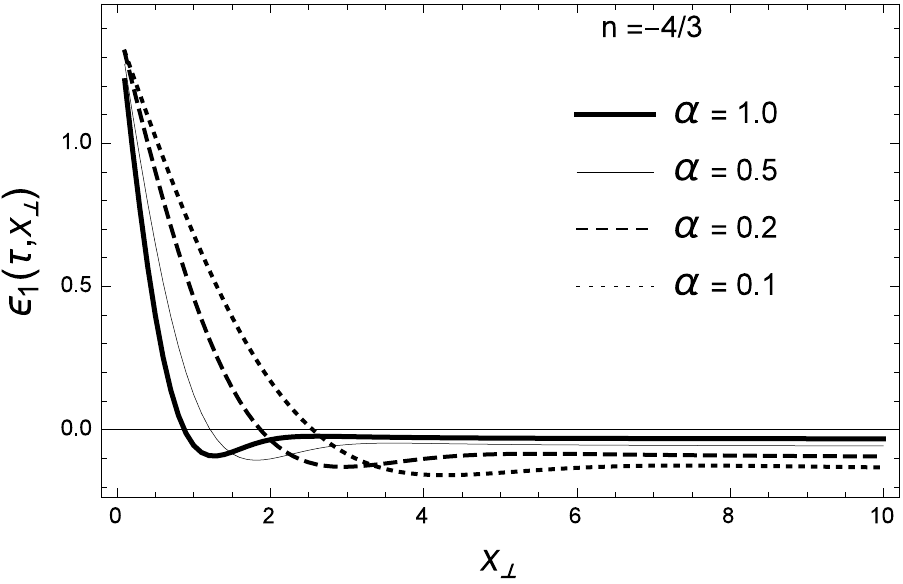} \\
\end{center}
\caption{$\epsilon_1(\tau,x_\perp)$ as a function of $x_\perp$ for different values of $\alpha$ at $\tau_0=1$~fm.}
\label{fig:figcom33}
\end{figure}

Also for the case $n=-4/3$, we plot the external magnetic profile at $\tau=1$~fm for several different
values of $\alpha$ in Fig. \ref{fig:figcom11}. In Figs. \ref{fig:figcom22} and \ref{fig:figcom33},
we plot $v_{x_\perp}$ and $\epsilon_1$  at $\tau=1$~fm. The qualitative behavior of $v_{x_\perp}$
and $\epsilon_1$ are similar to the case $n=-1$.
\\

\section{Transverse momentum spectrum in the presence of a weak external magnetic field}
In the previous sections we have obtained as analytical solution the transverse velocity and energy density
in the presence of a weak magnetic field. Now we can use these results to estimate the transverse momentum spectrum
emerging from the Magneto-hydrodynamic solutions.

From the local equilibrium hadron distribution the transverse
spectrum is calculated at the freeze out surface via the Cooper-Frye (CF) formula:
\begin{eqnarray}
S=E\frac{d^3N}{dp^3}=\frac{dN}{p_Tdp_T dyd\varphi}=\int d\Sigma_\mu p^\mu \exp(\frac{-p^\mu u_\mu}{T_f})
\end{eqnarray}
We note that $T_f$ is the temperature at the freeze out surface. The latter is the
isothermal surface in space-time at which the temperature of
inviscid  fluid is related to the energy density as $T\propto \epsilon^{1/4}$. It must satisfy $T(\tau, x_\perp)=T_f$.

In our convention,
\begin{eqnarray}
d\Sigma_\mu&=&(-1, R_f, 0, 0)\tau_f x_\perp dx_\perp d\varphi  d\eta,\\
p^\mu&=&(m_T\cosh(Y-\eta), P_T\cos(\varphi_p-\varphi), rp_T\cos(\varphi_p-\varphi), \tau_f m_T\sinh(Y-\eta)),\nonumber\\
d\Sigma_\mu p^\mu&=&[-m_T\cosh(Y-\eta)+p_T R_f\cos(\varphi_p-\varphi)]\tau_f x_\perp dx_\perp d\varphi d\eta,\\
p^\mu u_\mu&=&-m_T\cosh(Y-\eta)u_\tau+p_T\cos(\varphi_p-\varphi)u_\perp,
\end{eqnarray}
where $R_f\equiv -\frac{\partial \tau}{\partial x_\perp}=\frac{\partial_\perp T}{\partial_\tau T}\mid_{T_f}$.
Here $\tau=\sqrt{t^2-z^2}$ is the longitudinal proper time, $ x_\perp$ the transverse
(cylindrical) radius, $\eta=\frac{1}{2}\log\frac{t+z}{t-z}$ the longitudinal rapidity
(hyperbolic arc angle), the azimuthal angle $\varphi_p$ belonging to
the spacetime point $x^\mu$ . Similarly $u_\perp$ is the transverse flow velocity and $\varphi$ is its asimuthal angle.
Finally $p_T$ is the detected transverse momentum,  $m_T=\sqrt{m^2+p_T^2}$ the corresponding
transverse mass, while $Y$ is the observed longitudinal rapidity, which gives our final expression for the CF formula
\begin{eqnarray}
S=\frac{g_i}{2\pi^2}\int_0^{x_f}\  x_\perp\ \tau_f(x_\perp)\ dx_\perp\
\Big[m_T K_1(\frac{m_Tu_\tau}{T_f})I_0(\frac{m_Tu_\perp}{T_f})+p_T R_f
K_0(\frac{m_Tu_\tau}{T_f})I_1(\frac{m_Tu_\perp}{T_f})\Big]
\label{spectrum}
\end{eqnarray}
Where $\tau_f(x_\perp)$ is the solution of the $T(\tau_f, x_\perp)=T_f$ and the degeneracy is $g_i=2$
for both the pions and the protons. The above integral over $x_\perp$ on the freeze-out surface is evaluated numerically.

\begin{figure}[h!]
\begin{center}
\includegraphics[width=4in]{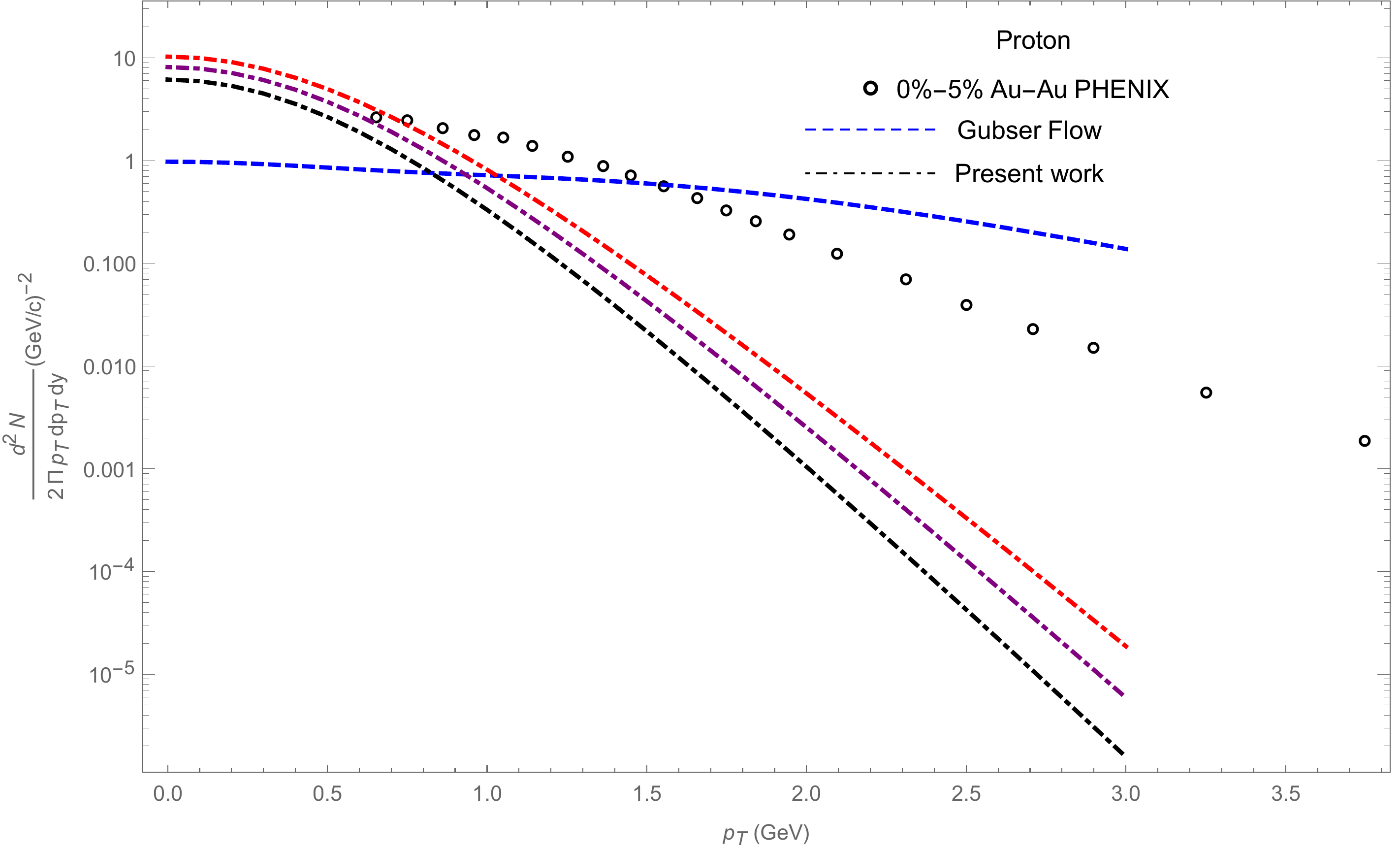} \\
\end{center}
\caption{Proton transverse spectrum from central Au-Au collisions: black, purple and red lines correspond to a
freeze out temperature of 140, 150 and 160 MeV, respectively. Circles: PHENIX data~\cite{phenix}.}
\label{fig:figproton}
\end{figure}

\begin{figure}[h!]
\begin{center}
\includegraphics[width=4in]{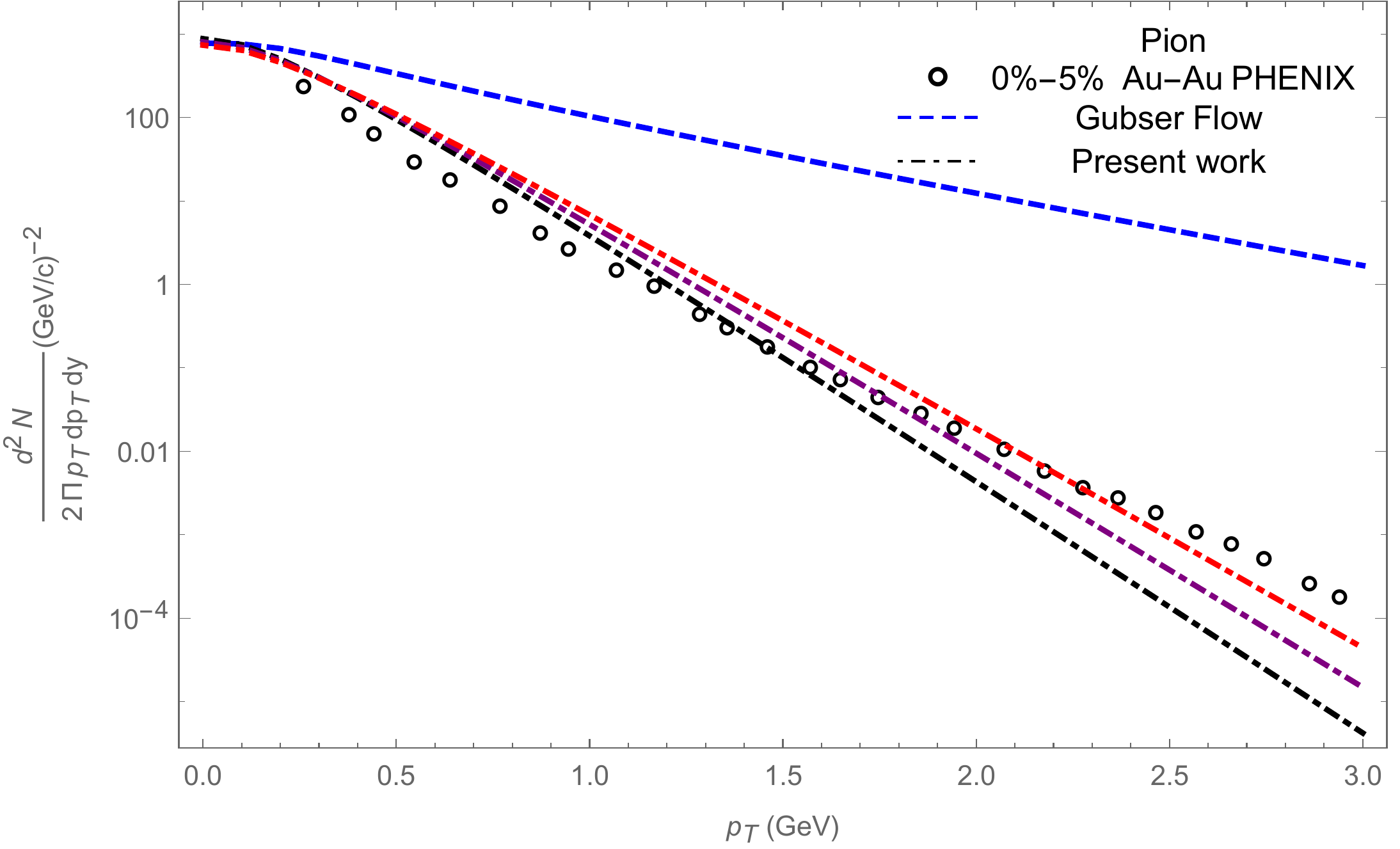} \\
\end{center}
\caption{Pion transverse spectrum from central Au-Au collisions: black, purple and red lines correspond to a
freeze out temperature of 140, 150 and 160 MeV, respectively. Circles: PHENIX data~\cite{phenix}.}
\label{fig:figpion}
\end{figure}

The spectrum Eq.~(\ref{spectrum}) is illustrated in Fig.~\ref{fig:figproton} and \ref{fig:figpion} for three different
values of the freeze out temperature (140, 150 and 160 MeV) and compared with experimental results obtained at PHENIX ~\cite{phenix}
in central collisions.
Our proton spectrum appear to underestimate the experimental data, except at low $p_T$, but their behavior with $p_T$has the correct trend of
a monotonical decrease. The pion spectrum, instead, appears in fair agreement with the experimental results, which are very close to the
theoretical curves. This is an indication that hadrons with different masses have different sensitivities to the underlying hydrodynamic flow
and to the electromagnetic fields. Indeed, the difference between the charge-dependent flow of light pions and heavy protons might arise
because the former are more affected by the weak magnetic field than the heavy protons.

For comparison, we also show the results obtained by Gubser, which appear to be more flat and typically overestimate the experiment.
We also notice that, for the proton case, the highest value of the freeze out temperature we employed (as suggested, e.g. in Ref.~\cite{Ratti})
slightly brings (for protons) the calculation closer to the experimental data; however it also shows a kind of saturation phenomenon
and points to the need of including other effects not considered in the present work.

\section{Conclusions}
In the present work, we investigated central heavy ion collisions in the presence of a transverse external
magnetic field. Making use of Milne coordinates, in our setup the medium is boost-invariant
along the z direction and the magnetic field, which is a function of $\tau$ and $x_\perp$, points along
the $\phi$ direction. The energy conservation and Euler equations reduced to two
coupled differential equations, which we solved analytically in the weak-field approximation.
We showed in detail how the fluid velocity and energy density are modified by the magnetic field.
The solutions obtained by our numerical calculations assume an initial energy density of the fluid at time $\tau=1$~fm
fixed to $\sim5.4$~ GeV/fm$^3$
and a ratio of the magnetic field energy to the fluid energy density, $\sigma$, fixed to $\sim0.015$.
We consider two different decays with time of the magnetic field: $\tau^n$, with $n=-1$ or $n=-4/3$.
A visual presentation of the flow for $n=-1$ can be find in Figs. \ref{fig:fig2} and \ref{fig:fig3} and
for $n=-4/3$ in Figs. \ref{fig:fig22} and \ref{fig:fig33}.

We remark that in Ref.~\cite{pu} the external magnetic
field was approximated by a Fourier cosine series and, due to the oscillatory behavior of the cosine function,
the magnetic field reduces to zero in the fringes for $|x|=\pi$. Consequently, these authors had to focus on
the valid region $-\pi<x<\pi$ and the behavior of the transverse velocity and of the correction energy
density was difficult to analyze near the fringes. In the present work the magnetic field is approximated
by a series of Bessel functions and the solutions are valid for the entire region of $x_\perp$,
i.e., $(0,\infty)$.

Another point concerning the choice of the $\tau$ dependence of the magnetic field is related
to the ratio $\sigma$ between magnetic and fluid energy densities: in Ref.~\cite{pu15}, it was found that
in central collisions, at the center
of the collision region, $\sigma\ll1$ for most of the events; nevertheless, large values of $\sigma$ were
observed in the outer regions of the collision zone.
Therefore, our assumption for the spatial distribution of the external
magnetic field for the case $n=-1$ may be more realistic at face of the physical conditions.

 In general, our study in a simple
setup, which includes an azimuthal magnetic field in the matter distribution, is worthwhile to check
the possible effect of this change on the transverse expansion of the fluid. We showed that by
combining the azimuthal magnetic field with the boost symmetry along the beam direction, a
radial flow perpendicular to the beam axis is created and  the energy density of the fluid is altered. We stress that the present
work presents an approximated calculation which can be useful for cross checking current and future numerical calculations in some limiting region.
Indeed, the effect of such a scenario on hadronic flow in heavy ion collisions requires more pragmatic debates.

Our study can be generalized in many directions: breaking of rotational symmetry can be introduced, the
conservation equation can be coupled to Maxwell's equations and solved consistently. Of course in this case
only numerical solutions can be found, while in the present paper we were able to obtain analytical solutions.

%*********************************************************************************
%************************************************************************************
%*********************************************************************************
%*************************************************


\begin{thebibliography}{99}

\bibitem{bj83} J. D. Bjorken, Phys. Rev. D 27, 140 (1983).

\bibitem{gubser} S. S. Gubser, " Symmetry constraints on generalizations of Bjorken flow ", Phys. Rev. D 82, 085027 (2010).

\bibitem{pu15} Victor Roy, Shi Pu, " Event-by-event distribution of magnetic field energy over
initial fluid energy density
in $\sqrt{S_{NN}}= 200$ GeV Au-Au collisions ", Phys. Rev C 92, 064902, (2015).

\bibitem{pu} Shi Pu, and Di-Lun Yang, '' Transverse flow induced by inhomogeneous magnetic fields
in the Bjorken expansion '', Phys. Rev D 93, 054042 (2016).

\bibitem{Tuchin:2013apa}
  K.~Tuchin,
  ``Time and space dependence of the electromagnetic field in relativistic heavy-ion collisions,''
  Phys.\ Rev.\ C {\bf 88} (2013) no.2,  024911.

\bibitem{Tuchin:2013ie}
  K.~Tuchin,
  ``Particle production in strong electromagnetic fields in relativistic heavy-ion collisions,''
  Adv.\ High Energy Phys.\  {\bf 2013} (2013) 490495.

\bibitem{a17} K. Tuchin, " Electromagnetic fields in high energy heavy-ion collisions ",
Int. J.  Mod. Phys. E Vol. 23, No. 1 (2014) 1430001.

\bibitem{a20} B. G. Zakharov, " Electromagnetic response of quark gluon plasma in heavy ion collisions ",
Phys. Lett. B 737 (2014) 262-266.
\bibitem{McLerran:2013hla}
  L.~McLerran and V.~Skokov,
  ``Comments About the Electromagnetic Field in Heavy-Ion Collisions,''
  Nucl.\ Phys.\ A {\bf 929} (2014) 184.

\bibitem{Deng:2012pc}
  W.~T.~Deng and X.~G.~Huang,
  ``Event-by-event generation of electromagnetic fields in heavy-ion collisions,''
  Phys.\ Rev.\ C {\bf 85} (2012) 044907.

\bibitem{chinees} H. Li, X-L. Sheng, Q. Wang, " Electromagnetic fields with electric and
chiral magnetic conductivities in heavy ion collisions ",
Phys. Rev.  C { \bf 94}, 044903 (2016)

\bibitem{Gursoy:2014aka}
  U.~Gursoy, D.~Kharzeev and K.~Rajagopal,
  ``Magnetohydrodynamics, charged currents and directed flow in heavy ion collisions,''
  Phys.\ Rev.\ C {\bf 89} (2014) no.5,  054905

\bibitem{Bzdak:2011yy}
  A.~Bzdak and V.~Skokov,
  ``Event-by-event fluctuations of magnetic and electric fields in heavy ion collisions,''
  Phys.\ Lett.\ B {\bf 710} (2012) 171.

\bibitem{a14} V. V. Skokov, A. Yu. Illarionov and V. D. Toneev ," Estimate of the magnetic field strength
in heavy-ion collision ",
Int. J.  Mod. Phys. A Vol. 24, No. 31 (2009) 5925–5932.

\bibitem{a15} Yang Zhong, Chun-Bin Yang, Xu Cai, and Sheng-Qin Feng,
" A Systematic Study of Magnetic Field in Relativistic Heavy-Ion Collisions in the RHIC and LHC
Energy Regions ",
Advances in High Energy Physics Volume 2014, Article ID 193039.

\bibitem{voronyuk} V. Voronyuk, V. D. Toneev, W. Cassing, E. L. Bratkovskaya, V. P. Konchakovski, S. A. Voloshin,"
 Electromagnetic field evolution in relativistic heavy-ion collisions ",
Phys. Rev C 83, 054911 (2011).

\bibitem{kharzeev2008} D. E. Kharzeev, L. D. McLerran, and H. J. Warringa, '' The effects of topological charge change in heavy ion collisions: "Event by event P and CP violation" '',
Nucl. Phys. A803, 227 (2008), arXiv:0711.0950 [hep-ph].

\bibitem{kharzeev2010} D. E. Kharzeev, '' Topologically induced local P and CP violation in QCD x QED '', Annals Phys. 325, 205 (2010),
arXiv:0911.3715 [hep-ph].

\bibitem{kharzeev2011} D. E. Kharzeev and H.-U. Yee, '' Chiral Magnetic Wave '', Phys. Rev. D83, 085007
(2011), arXiv:1012.6026 [hep-th].

\bibitem{kharzeevprl} Y. Burnier, D. E. Kharzeev, J. Liao, and H.-U. Yee, '' Chiral magnetic wave at finite baryon density and the electric quadrupole moment of quark-gluon plasma in heavy ion collisions  '',
Phys. Rev. Lett. 107, 052303 (2011), arXiv:1103.1307
[hep-ph].

\bibitem{roy15} V. Roy, S. Pu, L. Rezzolla, D. Rischke, " Analytic Bjorken flow in one-dimensional
relativistic magnetohydrodynamics ", Physics Letters B, Vol. 750, (2015).

\bibitem{pu16} S. Pu, V. Roy, L. Rezzolla and D. H. Rischke, " Bjorken flow in one-dimensional
relativistic magnetohydrodynamics with magnetization ", Phys. Rev. D 93, 074022 (2016).

\bibitem{pang16} L. G. Pang, G. Endr\"{o}di and H. Petersen, " Magnetic field-induced squeezing
effect at RHIC and at the LHC ", Phys. Rev. C 93, 044919 (2016).

 \bibitem{Inghirami etal} G. Inghirami, L. Del Zanna, A. Beraudo, M. Haddadi Moghaddam,
F. Becattini, M. Bleicher, " Numerical magneto-hydrodynamics for relativistic nuclear collisions ",
Eur. Phys. J. C (2016) 76:659.

\bibitem{Haddadi et al} M. H. Moghaddam, B. Azadegan, A. F. Kord, W. M. Alberico, " Non-relativistic approximate numerical
ideal-magneto-hydrodynamics of (1+1D) transverse flow in
Bjorken scenario ", Eur. Phys. J. C (2018) 78:255.

\bibitem{Das} A.~Das, S.S.~Dave, P.S.~Saumia and A. M. Srivastava, ``Effects of magnetic field on the
plasma evolution in relativistic heavy-ion collisions'', Phys. Rev. C 96, 034902 (2017).

\bibitem{Roy} V.~Roy, S.~Pu, L.~Rezzolla and D.H.~Rischke, ``Effects of intense magnetic fields on reduced-MHD
evolution in $\sqrt{s_{NN}}=200$~GeV Au+Au collisions'', Phys. Rev. C 96, 054909 (2017).

\bibitem{feng} B. Feng, Z. Wang, " Effect of an electromagnetic field on the spectra and elliptic flow of particles ", Phys. Rev. C 95, 054912 (2017).

\bibitem{greif} M. Greif, C. Greiner, Z. Xu, " Magnetic field influence on the early time dynamics of heavy-ion collisions ",  	Phys. Rev. C 96, 014903 (2017).

\bibitem{hattori} K. Hattori, Xu-G. Huang, D. Satow, D. H. Rischke, " Bulk Viscosity of Quark-Gluon Plasma in Strong Magnetic Fields ", Phys. Rev. D 96, 094009 (2017).

\bibitem{geod}J. Goedbloed, R. Keppens, S. Poedts"  Advanced Magnetohydrodynamics with Applications
to Laboratory and Astrophysical Plasmas ", Cambridge University Press, 2010.

\bibitem{an89} A. M. Anile, " Relativistic fluids and magneto-fluids ", Cambdridge University Press (1989).

\bibitem{phenix} K. Adcox et al. (PHENIX Collaboration),
``Formation of dense partonic matter in relativistic nucleus–nucleus collisions at RHIC: Experimental evaluation by the PHENIX Collaboration''
Nucl. Phys. A 757, 184 (2005)

\bibitem{Ratti}
  C.~Ratti, R.~Bellwied, J.~Noronha-Hostler, P.~Parotto, I.~Portillo Vazquez and J.~M.~Stafford,
  %``Freeze-out temperature from net-Kaon fluctuations at RHIC,''
  arXiv:1805.00088 [hep-ph].










\end{thebibliography}
\end{document}